\documentclass[prb,twocolumn,superscriptaddress,amsmath,amssymb,showpacs,floatfix,preprintnumbers]{revtex4-1}
\usepackage{amsmath}
\usepackage{amssymb}
\usepackage{graphicx}
\usepackage[export]{adjustbox}
\usepackage{epstopdf}
\usepackage{color}
\usepackage{epsfig}
\usepackage{braket}
\usepackage{amsmath}
\usepackage{stackengine}
\usepackage[colorlinks=true,citecolor=blue,linkcolor=blue]{hyperref}
\usepackage{amssymb}

\DeclareMathAlphabet{\bi}{OML}{cmm}{b}{it}

\begin{document}

\title{Molecular collapse in monolayer graphene}
\date{\today}
\author{R. Van Pottelberge}\email{robbe.vanpottelberge@uantwerpen.be}
\affiliation{Departement Fysica, Universiteit Antwerpen, Groenenborgerlaan 171, B-2020 Antwerpen, Belgium}

\author{D. Moldovan}
\affiliation{LUMICKS, De Boelelaan 1085, NL-1081 Amsterdam, The Netherlands}

\author{S. P. Milovanovi\'c}\email{slavisa.milovanovic@uantwerpen.be}
\affiliation{Departement Fysica, Universiteit Antwerpen, Groenenborgerlaan 171, B-2020 Antwerpen, Belgium}

\author{F. M. Peeters}\email{francois.peeters@uantwerpen.be}
\affiliation{Departement Fysica, Universiteit Antwerpen, Groenenborgerlaan 171, B-2020 Antwerpen, Belgium}
\begin{abstract}
Atomic collapse is a phenomenon inherent to relativistic quantum mechanics where electron states dive in the positron continuum for highly charged nuclei. This phenomenon was recently observed in graphene. Here we investigate a novel collapse phenomenon when multiple sub- and supercritical charges of equal strength are put close together as in a molecule. We construct a phase diagram which consists of three distinct regions: 1) subcritical, 2) frustrated atomic collapse, and 3) molecular collapse. We show that the single impurity atomic collapse resonances rearrange themselves to form molecular collapse resonances which exhibit a distinct quasi-bonding, anti-bonding and non-bonding character. Here we limit ourselves to systems consisting of two and three charges. We show that by tuning the distance between the charges and their strength a high degree of control over the molecular collapse resonances can be achieved.       
\end{abstract}

\maketitle
\section{Introduction}
In relativistic physics it was predicted that atomistic bound states could enter the positron continuum if the charge of the nucleus exceeded a certain critical value. After exceeding this critical charge the bound state hybdridizes with the positron continuum and turns into a resonant state [\onlinecite{Greiner}-\onlinecite{Zeldovich}]. This process causes a reconstruction of the Dirac vacuum and essentially makes the atom unstable, putting a natural limit on the periodic table [\onlinecite{Shytov}]. Despite several experimental attempts to confirm the existence of atomic collapse the results were not conclusive [\onlinecite{Schweppe}, \onlinecite{Cowan}]. For decades the detection of atomic collapse seemed elusive. 

The discovery of graphene, however, opened a new door for atomic collapse research. The relativistic nature of the charge carriers together with the fact that the critical charge for atomic collapse is significantly lower makes graphene an ideal platform to put the prediction of atomic collapse to the test [\onlinecite{Pereira}-\onlinecite{Shytov2}]. Indeed recently atomic collapse was for the first time observed in three different systems: (i) multiple charged $Ca$ dimers placed above a graphene sheet [\onlinecite{Crommie}], (ii) a vacancy created in the graphene lattice charged with an STM tip [\onlinecite{Peeters}], and iii) an induced potential in graphene using a sharp STM tip [\onlinecite{Peeters2}]. 

Next to showing the existence of atomic collapse in graphene, supercritical charges are also useful for spatial confinement of electrons in quasi-bound states. This makes atomic collapse resonances useful for the control and manipulation of charge carriers in graphene which is vital for the use of graphene in future electronic applications [\onlinecite{Neto}]. 

The observation of atomic collapse opened up the question: how will atomic collapse manifest itself in the presence of multiple charges and for different arrangements of charges? This will not only provide us with more insights in atomic collapse, it will also show if multiple charges could provide a platform to tune the behaviour of the atomic collapse resonances. 

To date atomic collapse in the presence of multiple charges has been the subject of theoretical studies both in gapped and gapless graphene [\onlinecite{Saffa}-\onlinecite{Egger}]. However, all these studies focused on charges that are individually subcritical. It was shown that by decreasing the distance between subcritical charges supercriticality could be achieved. Another system that has attracted attention was a dipole system consisting of two charges of equal strength but with opposite sign [\onlinecite{Gorbar}-\onlinecite{VanDuppen}].  

\begin{figure}
\includegraphics[scale=0.3]{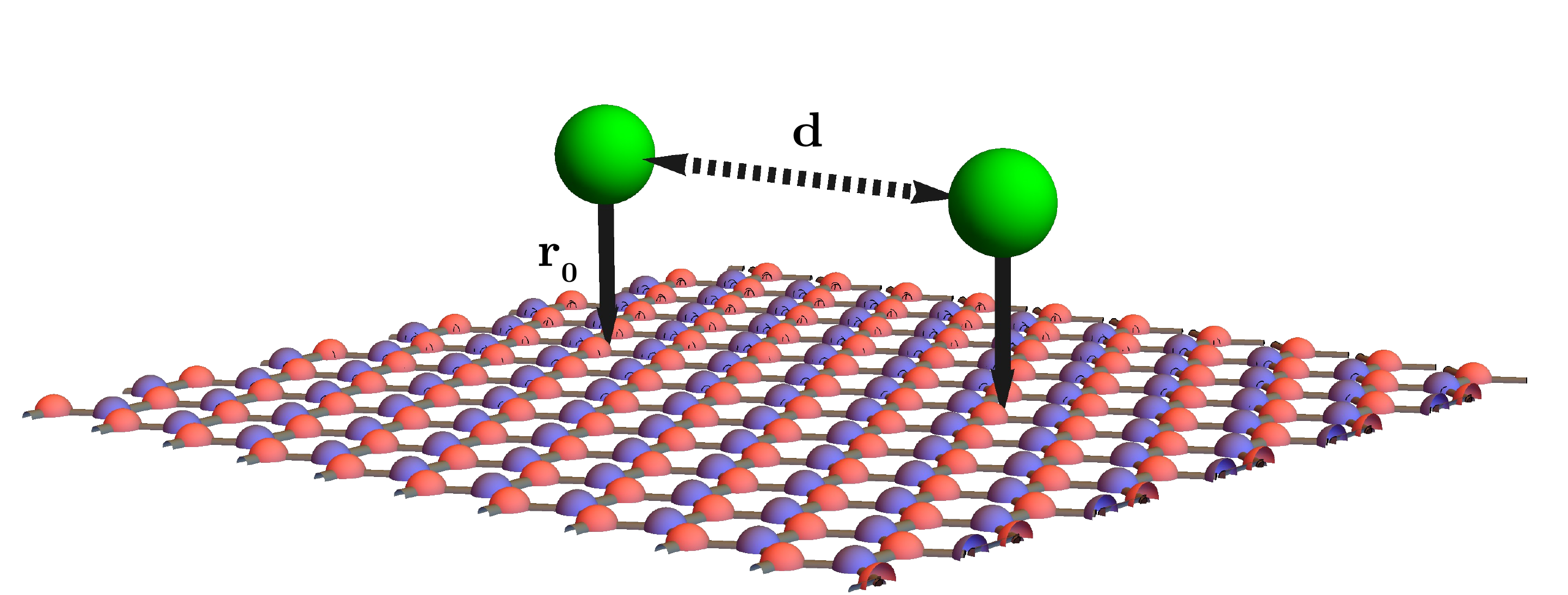}
\caption{Schematic representation of two equal charges (represented by green spheres) located at a distance $d$ from each other and a distance $r_0$ from the graphene lattice.}
\end{figure}

The focus of the present study is fundamentally different from previous investigations. We consider arbitrary values of the charges which are allowed to be individually supercritical and exhibit a corresponding single impurity atomic collapse resonance. We show that similar to how atomic orbitals in molecules form molecular orbitals, single impurity atomic collapse resonances of supercritical charges form molecular collapse resonances (i.e. quasi-bonding and anti-bonding states) which exhibit a spatial distribution which is fundamentally different from their single impurity counterparts. We show that these molecular collapse resonances can be tuned by their inter-charge distances and the size of the individual charges. Furthermore, we also consider the situation of three charges and investigate the differences with the two-charge problem.

The paper is structured as follows. In Sec. II we present the model. The main results for a two-charge system are presented in Sec. III. In Sec. IV we extend our results to a system consisting of three charges. The main conclusions of this work are presented in Sec. V.  
\section{Model}
Here, the tight-binding model will be used which in contrast to the continuum model is not limited to energies close to the Dirac point. For graphene the tight-binding Hamiltonian is given by the following expression [\onlinecite{Neto}]:
\begin{align}
\begin{split}
\hat{H}=\sum_{\langle i,j\rangle}\left(t_{ij}a_i^{\dagger}b_j+H.c.\right) \\ + \sum_i V(\vec{x}_i^A,\vec{y}_i^A)a_i^{\dagger}a_i+\sum_{i}V(\vec{x}_i^B,\vec{y}_i^B)b_i^\dagger b_i.
\end{split}
\end{align}
Here the first term represents the tight-binding Hamiltonian without any external fields. The hopping parameter is given by $t_{ij}$ and for graphene we take the generally accepted value $-2.8$ eV for nearest neighbour hopping. The operators $a_i(a_i^\dagger)$ and $b_i(b_i^\dagger)$ create(annihilate) an electron on the $i^{\text{th}}$ site of sublattice $A$ and $B$, respectively. The last two terms include an arbitrary electrostatic potential which for our case is due to the presence of Coulomb charges. $\vec{x,y}_i^{A/B}$ is the position of the carbon atoms.

The singularity of the point size Coulomb potential makes the problem ill-defined in the supercritical regime. In the latter case it is essential to perform a regularisation of the Coulomb potential [\onlinecite{Zarenia}] which is naturally present in any experimental set-up. In this paper we opt for the following regularized Coulomb potential:
\begin{equation}
V(x-x_0,y-y_0)=-\hbar v_F\frac{\beta}{\sqrt{(x-x_0)^2+(y-y_0)^2+r_0^2}}.
\end{equation} 
Here $(x_0,y_0)$ denote the cartesian coordinates of the position of the charge, $\beta=Z\alpha$ is the fine structure constant ($\alpha$) multiplied with the value of the charge ($Z$). For individual charges one has for $\beta<0.5$ the \textit{subcritical} regime and for $\beta>0.5$ the \textit{supercritical} regime. $r_0$ is the regularization distance, that is determined by the particular experimental set-up. The above potential corresponds to the potential felt by charge carriers in graphene due to a charge placed at a distance $r_0$ of the graphene sheet, see Fig. 1 [\onlinecite{Mayer}]. A reasonable value is $r_0=0.5$ nm which we will use throughout the present paper and which is in line with experimental data [\onlinecite{Crommie}, \onlinecite{Peeters}]. Previous experimental resuts could be explained using such a Coulomb potential. We recongnize that there are other types of regularizations possible [\onlinecite{Zhu}, \onlinecite{Gusynin}], which however will not have any influence on the essential physics that will be discussed in the present paper.  

Regarding the computation, a large hexagonal flake (edge size of $200$ nm which involves four million carbon atoms) was constructed. By placing the charges in the middle of the flake and due to the large size of the latter the physics will not be influenced by finite size effects. The LDOS will be calculated at the position of the Coulomb charge in the graphene lattice. For the calculation of the local density of states (LDOS) and spatial LDOS we used the open source tight binding package Pybinding [\onlinecite{Moldovan2}] which uses a kernel polynomial expansion to calculate the LDOS. 

\section{Two-charge system}
The most straightforward system of multiple charges is obviously the one consisting of two charges. This system is formally equivalent to the $H_2$ molecule which is often used as a seminal example for the linear combination of orbitals method (LCAO-method) [\onlinecite{Heitler}-\onlinecite{Mulliken}]. It is well known that the two $1S$ orbitals of both hydrogen atoms overlap with each other and hybridize forming two molecular orbitals, a higher in energy \textit{anti-bonding} orbital and lower in energy \textit{bonding} orbital.  
As mentioned in the introduction of this paper the atomic collapse resonances for single impurities resemble atomic orbitals and are essentially their unstable counterparts. We will show that due to this resemblance molecular bonding and anti-bonding resonances can be created by bringing two supercritical charges closer to each other. 

\begin{figure}
\includegraphics[scale=0.48]{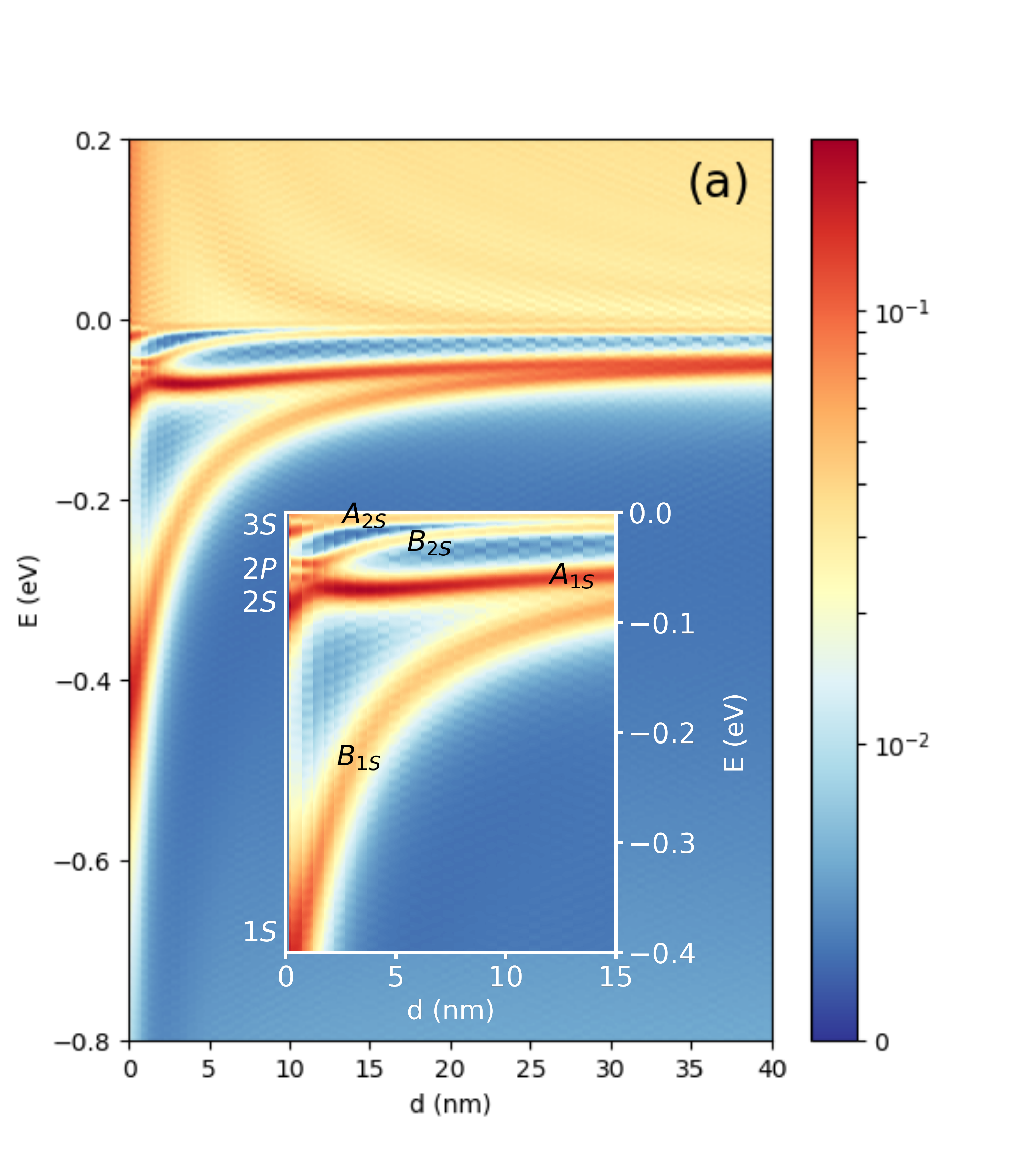}
\includegraphics[scale=0.48]{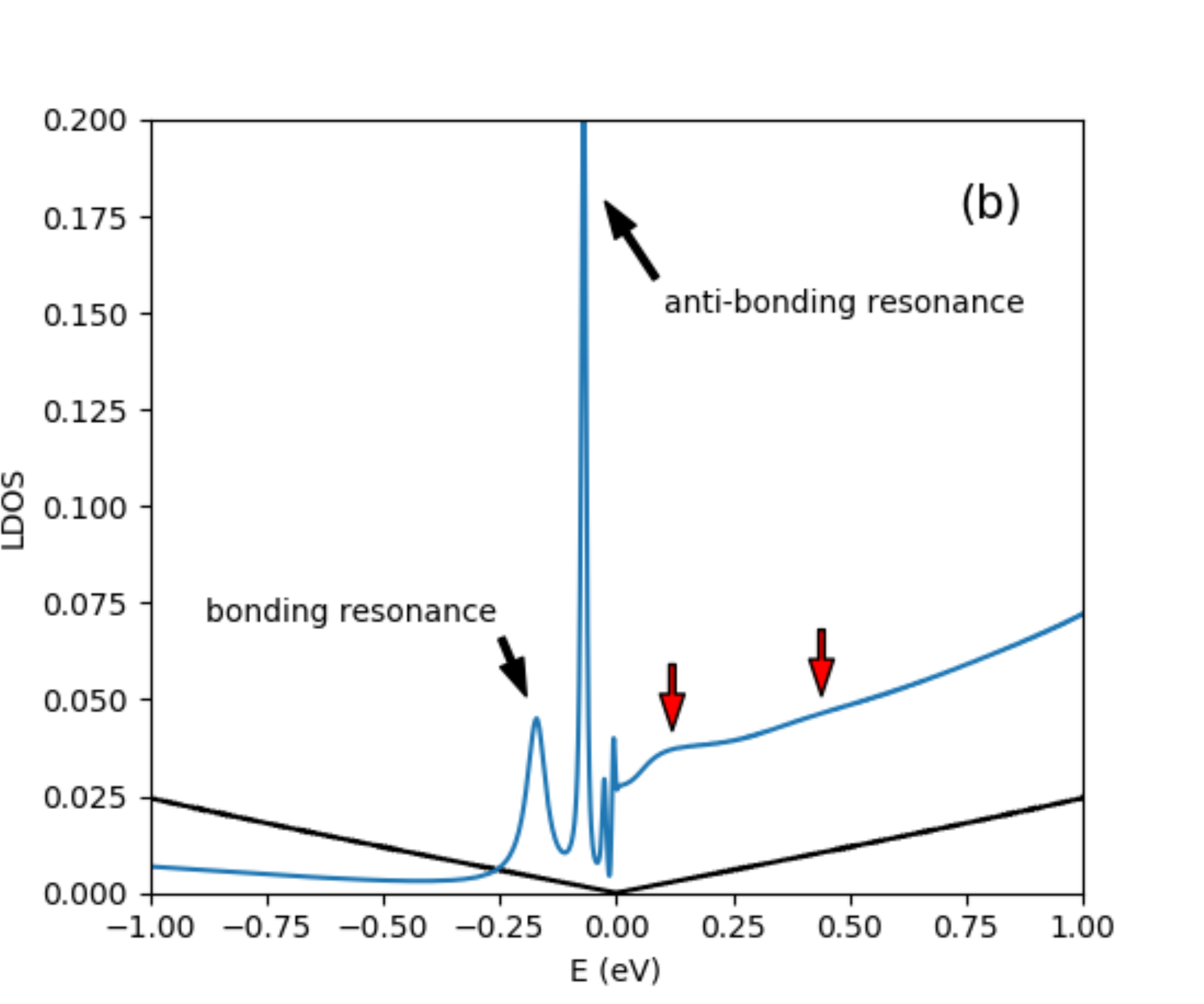}
\caption{(a) Density plot (log scale is used) of the LDOS at the position of one of the impurities in the graphene lattice as function of the inter-charge distance and energy. The inset shows a zoom of the energy range just below the Dirac point and for small inter-impurity distances. The different molecular collapse resonances are marked by black labels while the single impurity atomic collaps resonances are marked by white labels. An energy broadening of $0.003$ eV was used. Two individually supercritical charges with $\beta=1$ were used. (b) Cut of the top figure for two charges at a fixed distance $d=5$ nm from each other. The LDOS of pristine graphene is shown by the black curve. In this calculation a slightly larger flake size of 300 nm was used. In all calculations we used $r_0=0.5$ nm.}
\end{figure}

\begin{figure}
\includegraphics[scale=0.45]{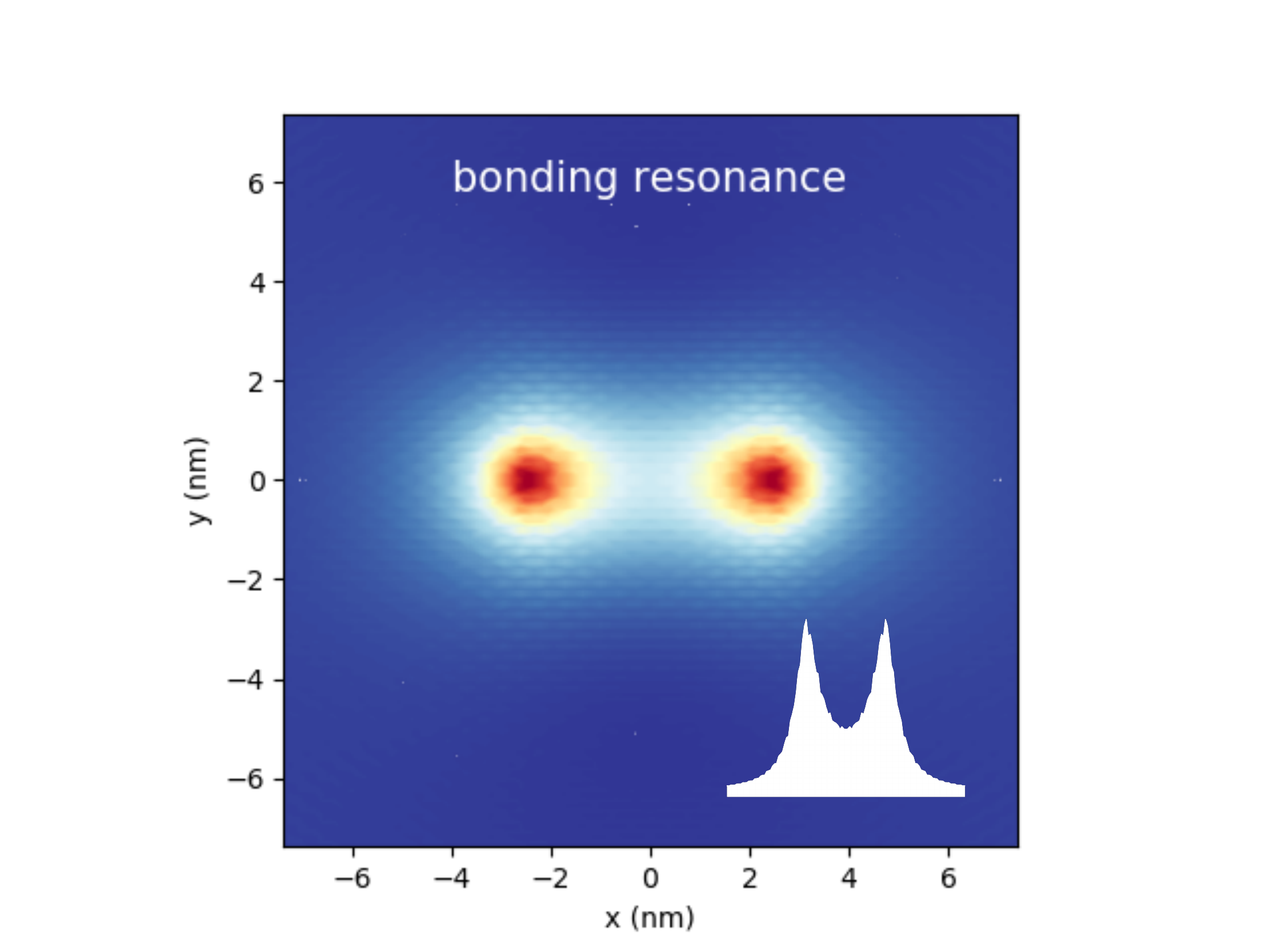}
\includegraphics[scale=0.45]{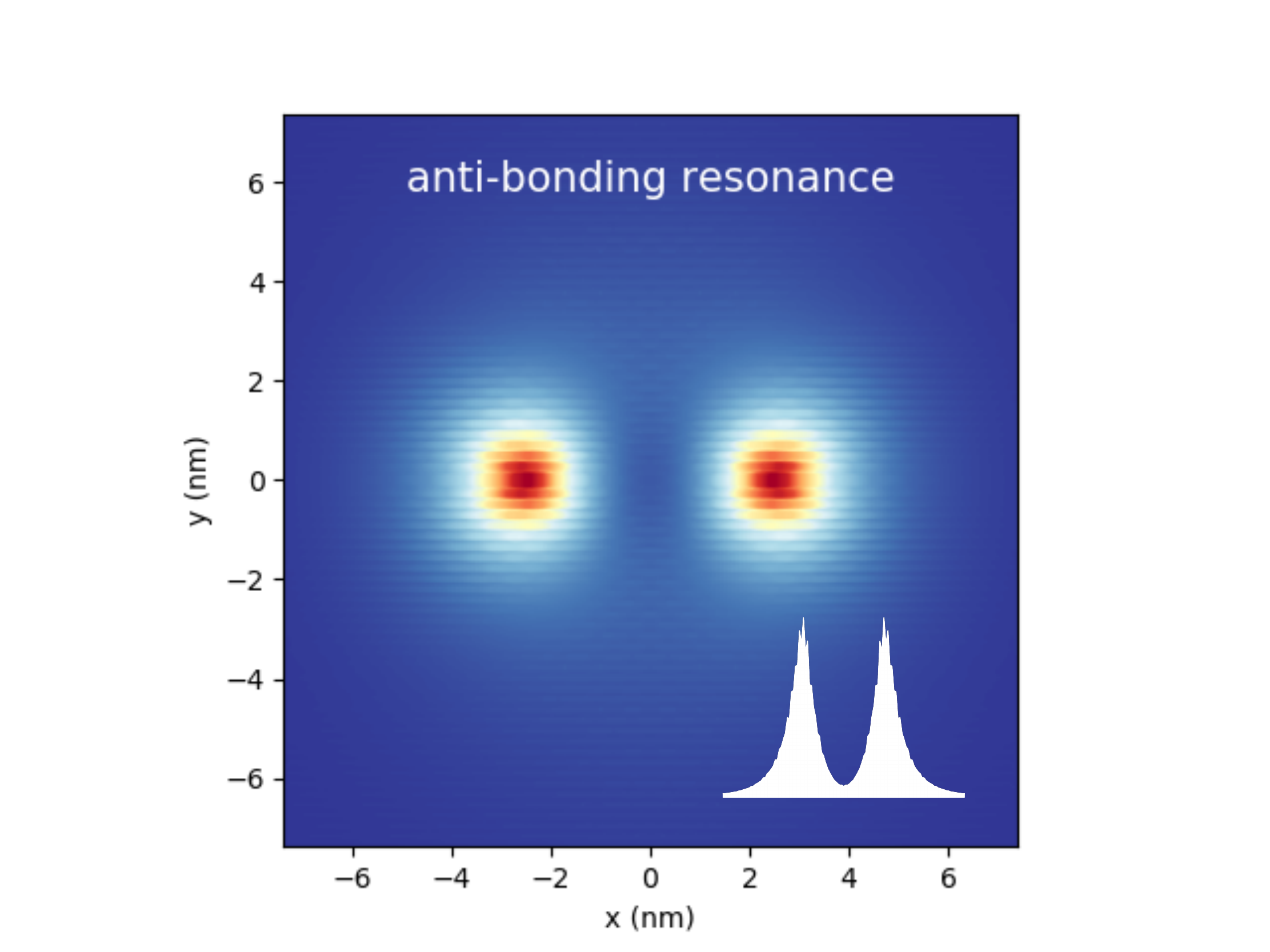}
\caption{Contour plot of the spatial LDOS for the bonding (top figure) and anti-bonding (bottom figure) resonance for two impurities separated at a distance $d=5$ nm from each other and individual charge $\beta=1$. The spatial LDOS is calculated for the energies $E=-0.16$ eV (bonding: $B_{1S}$) and $E=-0.08$ eV (anti-bonding: $A_{1S}$), respectively. A linear color scale is used with red corresponding to high LDOS and blue corresponding to zero LDOS. The inset figures in white show a cut of the spatial LDOS along $y=0$.}
\centering
\end{figure}


In Fig 2. we show the LDOS calculated within the tight-binding method for two Coulomb charges separated at a distance $d$ from each other, modelled by the potential (see schematic representation in Fig. 1):
\begin{equation}
\begin{split}
V_2(x,y)=-V(x-d/2,y)-V(x+d/2,y).
\end{split}
\end{equation}
The LDOS is shown as function of the inter-charge distance and energy. Two individually supercritical charges with a dimensionless charge of $\beta=1$ were used in the calculation. The LDOS itself is calculated at one of the impurity sites. Due to symmetry, probing one or the other impurity will produce the same result for the LDOS. 

\begin{figure}
\includegraphics[scale=0.45]{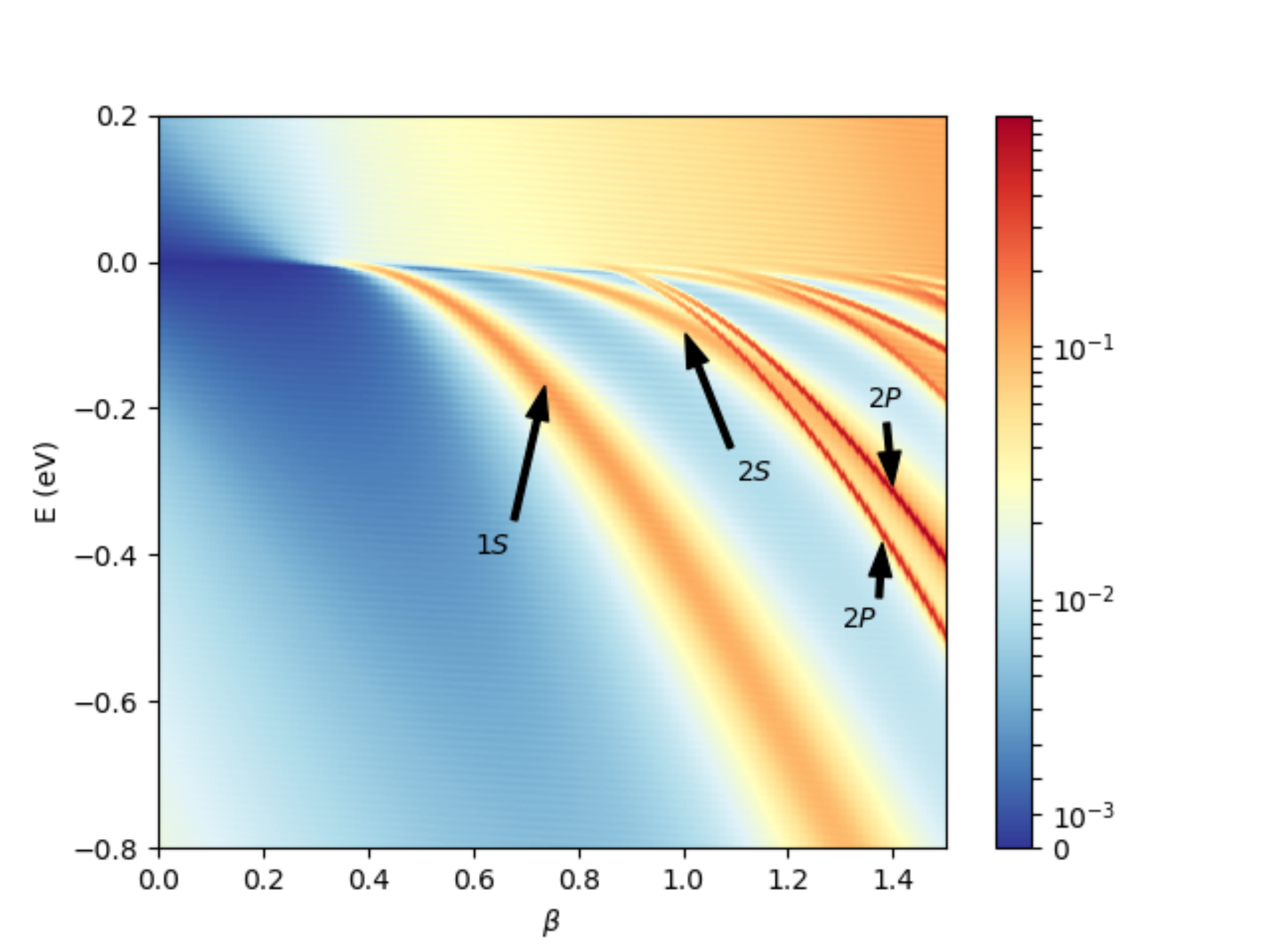}
\caption{LDOS for two charges at a distance $d=0$ from each other. The individual charge is $\beta$ and this situation corresponds to a single impurity with charge $2\beta$. The single impurity resonances are marked by $1S$, $2S$ and $2P$.}
\end{figure}

\begin{figure*}
\includegraphics[trim={0cm 0cm 0cm 0cm},scale=0.22]{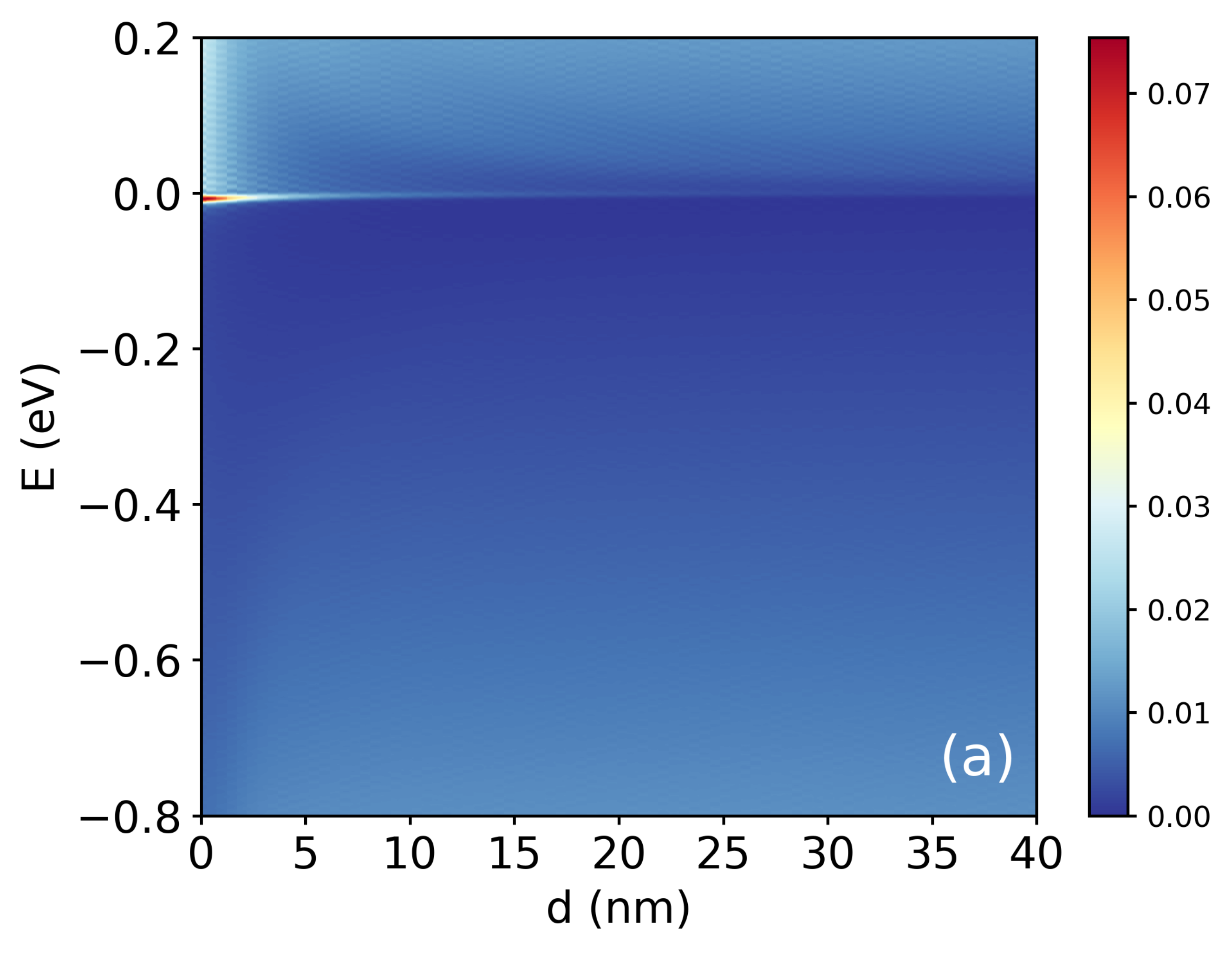}
\includegraphics[trim={0cm 0cm 0cm 0cm},scale=0.22]{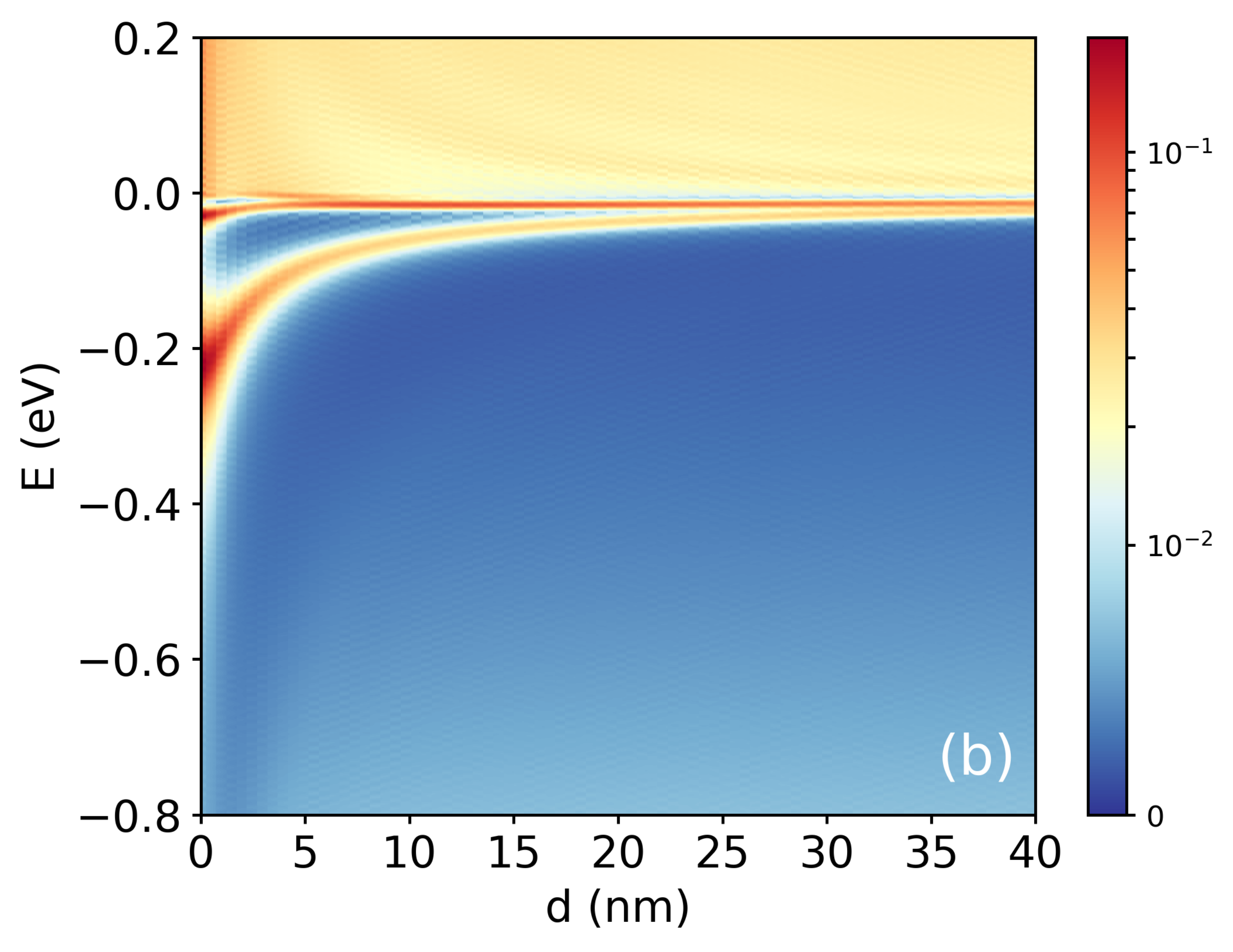}
\includegraphics[trim={0cm 0cm 2cm 0cm},scale=0.22]{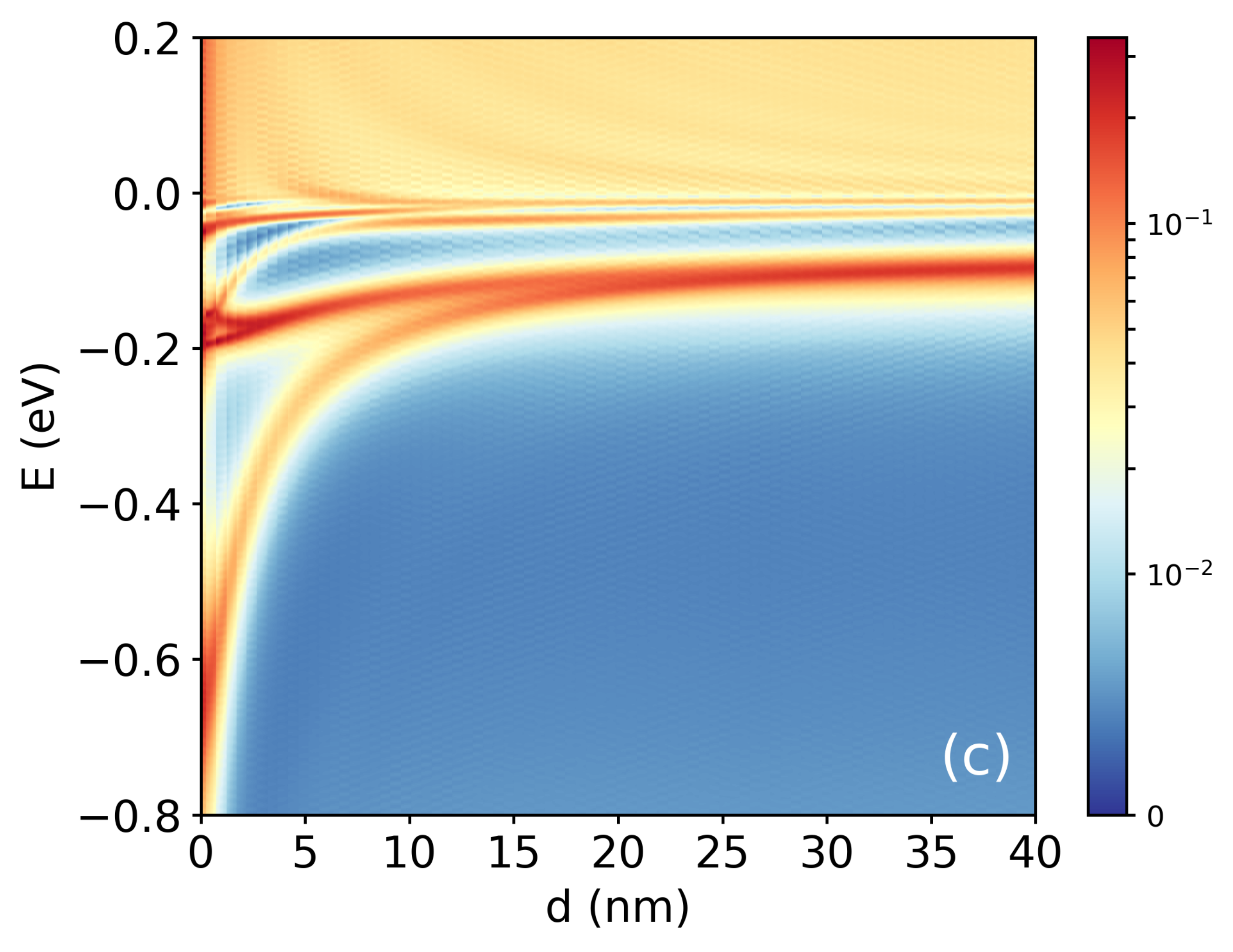}
\caption{Countour plot of the LDOS measured at one of the charges as function of the distance between the two charges for three different values of the two individual charges: (a) subcritical charges $\beta=0.4$, (b) supercritical charges with $\beta=0.8$ and (c) supercritical charges with $\beta=1.2$.}
\end{figure*}

In Fig. 2(a) we clearly observe one peak in the LDOS for large inter-charge distance. This corresponds to the atomic collapse resonance of a single Coulomb charge with $\beta=1$. In the literature this resonance is often labeled as the $1S$ resonance, which comes from the fact that it is the unstable counterpart of the $1S$ atomic orbital. When the inter-charge distance is decreased the single impurity resonance starts to split into two new resonances which are distinctly visible in the hole continuum. One branch decreases in energy (labeled as $B_{1S}$) while the other remains more or less constant in energy (labeled as $A_{1S}$). Decreasing the inter-charge distance even further we see that the lower branch starts to show a more profound dependence on the energy with decreasing inter-charge distance. The upper branch has a higher intensity, when measured on one of the impurity sites, compared to the lower branch. These observations show that the two resonances are quite different from the single impurity atomic collapse resonances. For postive energy broader less visible resonances are observed which decrease with increasing inter-charge distance. These resonances are for example observed in Fig. 2(a) for $E>0$ and in Fig. 2(b) marked by red arrows. These resonances originate from interference of electron scattering states between the two charges. The origin of these resonances is discussed in appendix A. 

In order to understand the behavior of these new states we show the spatial distribution of the LDOS in Fig. 3 when the two charges are at a distance $d=5$ nm from each other. For the lower energy branch (top figure) one can clearly see that the spatial LDOS has a finite density between the two impurities which strengthens the bond between them (also visible in the white inset figure which shows a cut along $y=0$). This behaviour, is typical for a molecular bonding orbital, and therefore this resonance is its unstable counterpart. The higher energy branch has a spatial LDOS that is zero between the two impurities weakening the bond between them, a behaviour that is characteristic for an anti-bonding molecular orbital. In the LCAO approximation the bonding and anti-bonding molecular orbitals are given by respectively 
\begin{equation}
\ket{B_{1S}}=\sqrt{\frac{1}{2}}\left(\ket{1S_a}+\ket{1S_b}\right),
\end{equation}
and
\begin{equation}
\ket{A_{1S}}=\sqrt{\frac{1}{2}}\left(\ket{1S_a}-\ket{1S_b}\right). 
\end{equation}
With $\ket{1S_{a}}$ and $\ket{1S_{b}}$ the single impurity orbitals on the separate impurities. From Fig. 3 we see the correspondence between the molecular collapse resonances and the LCAO method for atomic orbitals confirming that these resonances are indeed the unstable counterparts of molecular orbitals. 

\begin{figure}
\includegraphics[trim={1cm 0 1.5cm 0}, scale=0.50]{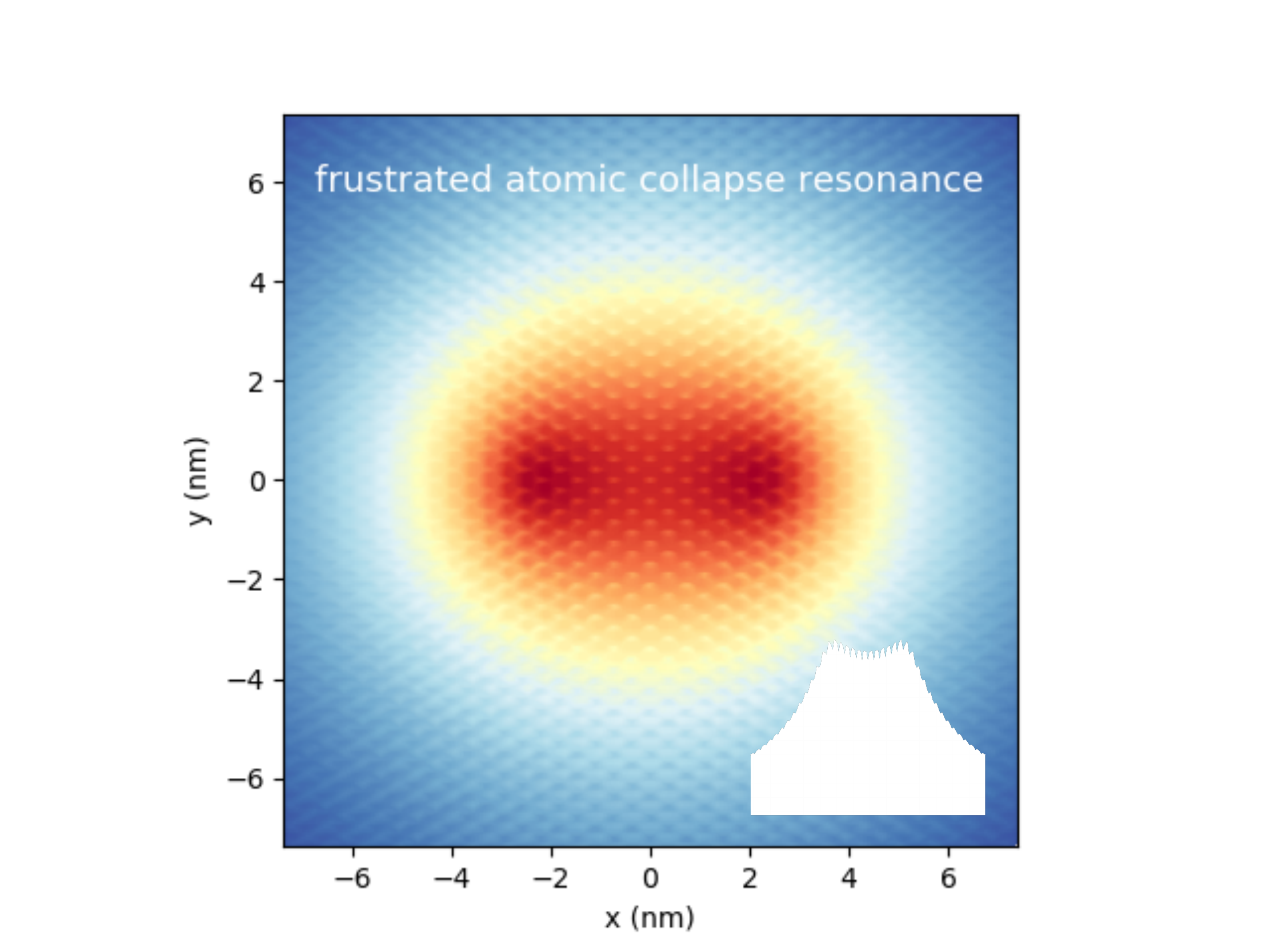}
\caption{Countour plot of the spatial LDOS for the frustrated atomic collapse resonance seen in Fig. 5(a). The spatial LDOS is calculated for $E=-0.004$ eV, $\beta=0.4$ and two charges seperated at a distance $d=5$ nm from each other. A linear scale was used with red representing high LDOS and dark blue zero LDOS.}
\end{figure}

For low d-values, one notices above the two branches another series of bonding and anti-bonding resonances right below the Dirac point, labeled respectively as $B_{2S}$ and $A_{2S}$. These two branches correspond to the splitting of the $2S$ single impurity resonance which is the unstable counterpart of the $2S$ atomic collapse state. In the inset of Fig. 2(a) a close up of the split branches is shown, the splitting of the $2S$ single impurity resonance can be clearly seen above the more profound splitting of the $1S$ resonance discussed in the previous paragraph. Note that for small $d$ their seems to be an increase in the number of resonances. For small $d$ above the broader resonance two smaller peaks appear, which can be clearly seen in the inset of Fig. 2(a) (left op the white $2P$ label). These two smaller peaks are related to higher angular momentum resonances that should appear in the single impurity ($\beta=2$) case. This is shown in Fig. 4 where the spectrum of two charges is shown for $d=0$ as function of the impurity strength $\beta$. For $\beta=1$ two narrower peaks are seen above the broader $2S$ resonance peak. These peaks are labeled with $2P$ since they can be considered as the unstable counterparts of the atomistic $2P$ state. Note that these extra peaks disappear with increasing inter-charge distance which is expected since two single charges with $\beta=1$ should not exhibit higher angular momentum resonances. We marked the single impurity resonances that appear when $d=0$ in Fig. 2(a) with white labelling using the same convention as in Fig. 4. The width of these resonances are mostly influenced by: i) the energy of the state, the larger the energy the larger the width of the resonances, this can be seen in Fig. 4 where the width clearly increases with increasing energy. ii) the angular momentum of the states also determines the width, the higher the angular momentum the smaller the width, this can be seen in Fig. 4 where the P states have clearly a smaller width compared to the S states.

In the bottom figure of Fig. 2(b) the LDOS is shown as function of the energy for two charges at a fixed inter-charge distance $d=5$ nm. As reference the LDOS of pristine graphene is shown by the black curve. The first molecular bonding and anti-bonding resonances should be clearly visible in experimental measurements, especially the profound difference in strength of both peaks should be a distinct signature to look for in experiments. The peaks just below the Dirac point are a result of the emergence of higher resonant states. 


In order to investigate the different collapsing regimes we study the dependence of the spectrum on the value of the charge $\beta$. This is shown in Fig. 5 where the LDOS as function of the inter-charge distance is shown for three different values of the individual charges: $\beta=0.4,\text{ }0.8, \text{and } 1.2$.

When $\beta=0.4$, see Fig. 5(a), both charges are individually subcritical and do not support a resonant state individually since it is less than the single impurity critical charge threshold $\beta_c=0.5$. However, when the charges are pushed close enough to each other their collective effect enables the emergence of a single weak resonance just below the Dirac point. This collective effect of individually subcritical charges has been studied theoretically both in gapped and gapless graphene [\onlinecite{Saffa}-\onlinecite{Egger}] and has been recently observed experimentally for an array of subcritical charges [\onlinecite{Crommie1}]. The regime where two charges are individually subcritical but together form a supercritical charge cluster was called \textit{frustrated atomic collapse} [\onlinecite{Crommie1}]. In this regime the localisation of the resonance is determined by the inter-charge distance $d$. As the inter-charge distance increases the resonance gets spread out over larger distances and becomes quenched as observed in Fig. 5(a). The spatial dependence of the frustrated atomic collapse is very different from the spatial distribution of the molecular collapse resonances. This is shown in Fig. 6 where we plot the spatial LDOS for the resonance observed in Fig. 5(a) for $E=-4$ meV and $d=5$ nm. The spatial LDOS is distributed over a larger area as compared to the densities shown in Fig. 3. This behaviour can be explained as follows. The frustrated atomic collaps regime can only occur when the charges are individually subcritical but together surpass the supercritical treshold, i.e. $\beta>0.5$. In that case charge carriers in graphene far away from the impurity charges will effectively feel one supercritical charge, hence they feel a single impurity potential $V(r)=-\beta/r$ with $\beta>0.5$. From the single impurity atomic collapse results we know that such a potential is able to induce atomic collapse. Consequently a resonance will emerge. However unlike the single impurity case where the localisation length scale is determined by the regularisation cutoff $r_0$, here for the frustrated atomic collapse it is the inter-charge distance $d$ that gives the length scale for the localisation. In short one can conclude that the physics of the frustrated atomic collapse regime is similar to the physics of the supercritical single impurity regime but with the regularisation paramter $r_0$ replaced by the much larger inter-charge distance $d$ explaining both the spatial distribution and strength of the frustrated atomic collapse resonance.     


The physics is totally different when both charges become individually supercritical, which is visible in Fig. 5(b) were we plotted the LDOS for $\beta=0.8$. In that case resonances exist for all values of the inter-charge distance. The single impurity resonance visible at larger inter-impurity distances splits up in a bonding and anti-bonding molecular resonance with a lower and higher LDOS, respectively. The resonances are located at higher energies compared to those shown in Fig. 2 which is a consequence of the smaller value of the individual charges. 

If we further increase the strength of the individual charges (Fig. 5(c), $\beta=1.2$) we see that the bonding and anti-bonding molecular resonances are shifted to lower energies. The splitting of the atomic collapse resonance into the molecular collapse resonances occurs at smaller inter-charge distance. The next splitting corresponding to the next bonding and anti-bonding molecular resonances which is now even more clearly visible as compared to the results shown in Fig. 2. 

Note that for $\beta=0.4$ and $\beta=0.8$ the number of resonances remains the same for all inter-charge distances while for $\beta=1.2$ a splitting leads to two additional resonances for smaller inter-charge distance. This can again be explained by the fact that for $d=0$ two additional resonances related to higher angular momentum states should appear which become quenched for increasing inter charge-distance.

In Fig. 7 we show the dependence of the spectrum on the impurity charge $\beta$ for the two charges seperated at a fixed distance $d=10$ nm from each other. For charges $\beta<0.25$ no atomic collapse resonance is visible below the Dirac point. When the charge is increased to $\beta\approx 0.25$ a resonance emerges below the Dirac point. However, since both charges are individually subcritical the resonance emerges due to the collective effect and we have entered the frustrated atomic collapse regime. If the charge is further increased beyond $\beta=0.5$ we know that individually both impurities support an atomic collapse resonance. At this point we have entered the new molecular collapse regime which is marked by the appearance of the anti-bonding resonance shortly after $\beta\approx 0.6$. Increasing the charge even further the bonding and anti-bonding resonances merge into one single resonance. This can be explained from the fact that with increasing charge the quasi-bound states become more localised around the individual charges, consequently behaving like single impurity resonances.

Now that we have studied the quasi-bound state spectrum for different values of the indivual charges and inter-charge distances we can capture the physics governing the system in a phase diagram. This is done in Fig. 8 where the three distinct regimes are marked in the phase diagram:

1) \textit{Subcritical region}: in this region each of the charges are individually subcritical and consequently do not support an atomic collapse resonance. Regardless of the distance between the two charges a resonance is not observed.

2) \textit{Frustrated atomic collapse region}: in this region both charges individually are subcritical but together form a supercritical charge for sufficient small $d$. This occurs when $\beta>0.25$ since the critical charge for the single impurity case is $\beta=0.5$. In this region resonances are possible for sufficient small $d$ due to the collective effect of the charges as recently observed experimentally in Ref. [\onlinecite{Crommie1}]. In this regime charge carriers in the far field region will feel a supercritical Coulomb potential. As a consequence as in the single impurity case an atomic collapse resonance is possible. However, the spatial distribution of this resonance is not determined by the regularisation parameter $r_0$ but by the much larger inter-charge distance $d$. Consequently the resonance becomes spread out with increasing inter-charge distance and quenches. This is indicated by the dotted line in Fig. 8 which is the point for which the LDOS at the impurity site for the lowest resonance is below the value $0.01$. However, it should be mentioned that this boundary is not sharp and only illustrative since the quenching occurs smoothly as seen in Fig. 5.

\begin{figure}[h!!!]
\includegraphics[scale=0.45]{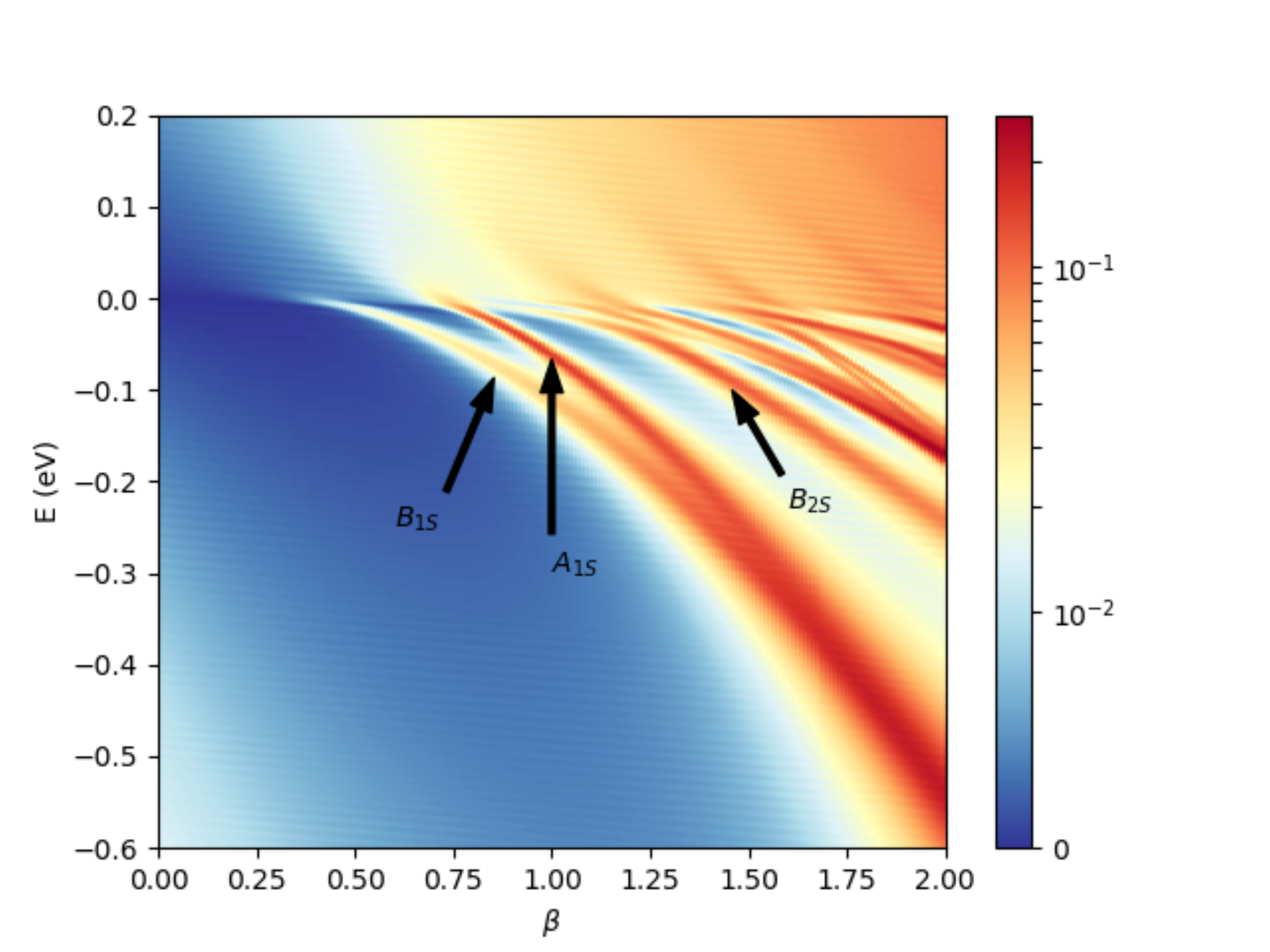}
\caption{Contour plot of the LDOS measured at one of the impurity sites as function of the impurity strength for two charges seperated by a distance $d=10$ nm from each other.}
\end{figure}

\begin{figure}[h!!!]
\includegraphics[scale=0.65]{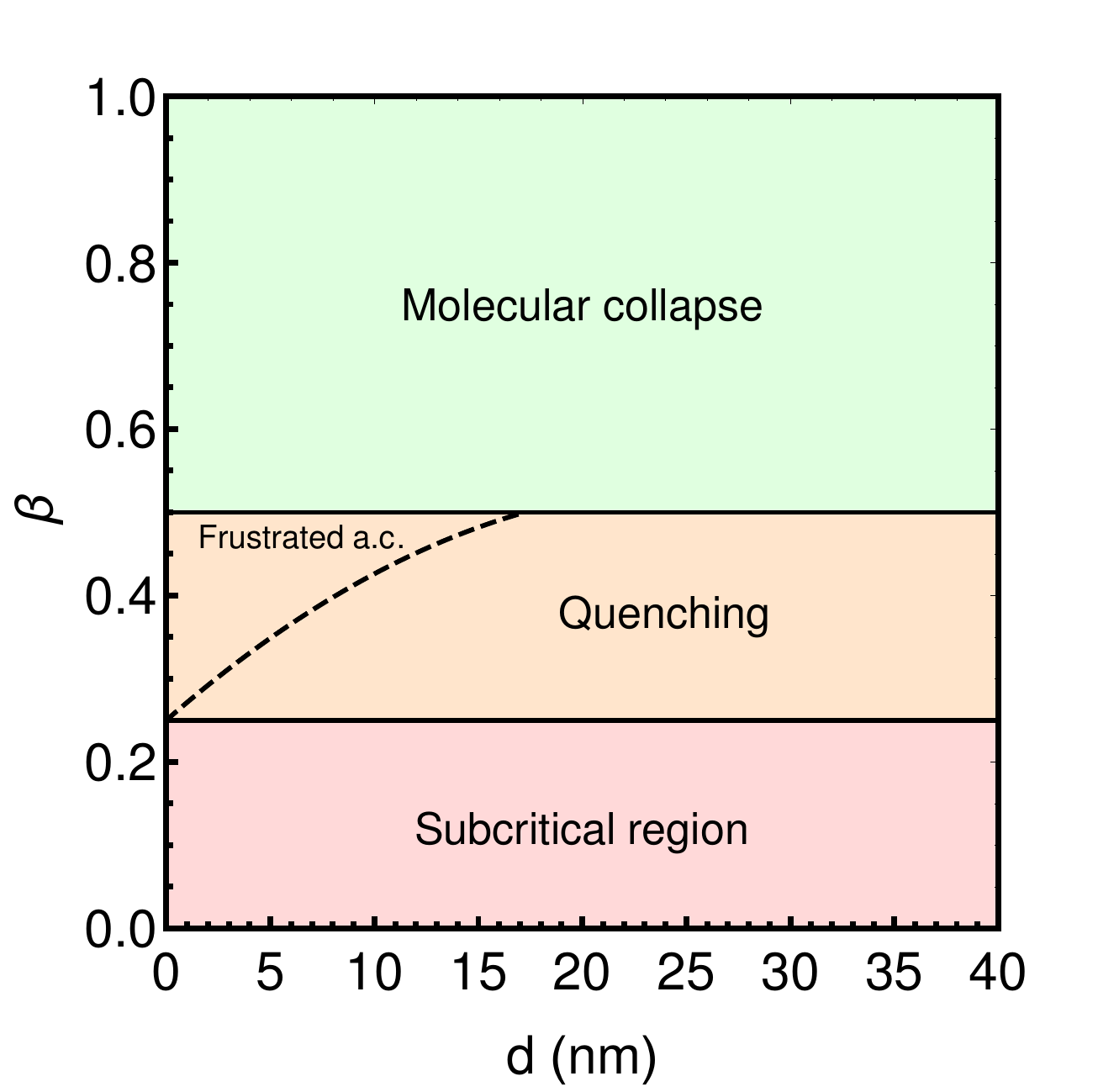}
\centering
\caption{Phase diagram of the different molecular collapse regimes for a two-charge system.}
\end{figure}


\begin{figure*}
\includegraphics[scale=0.47]{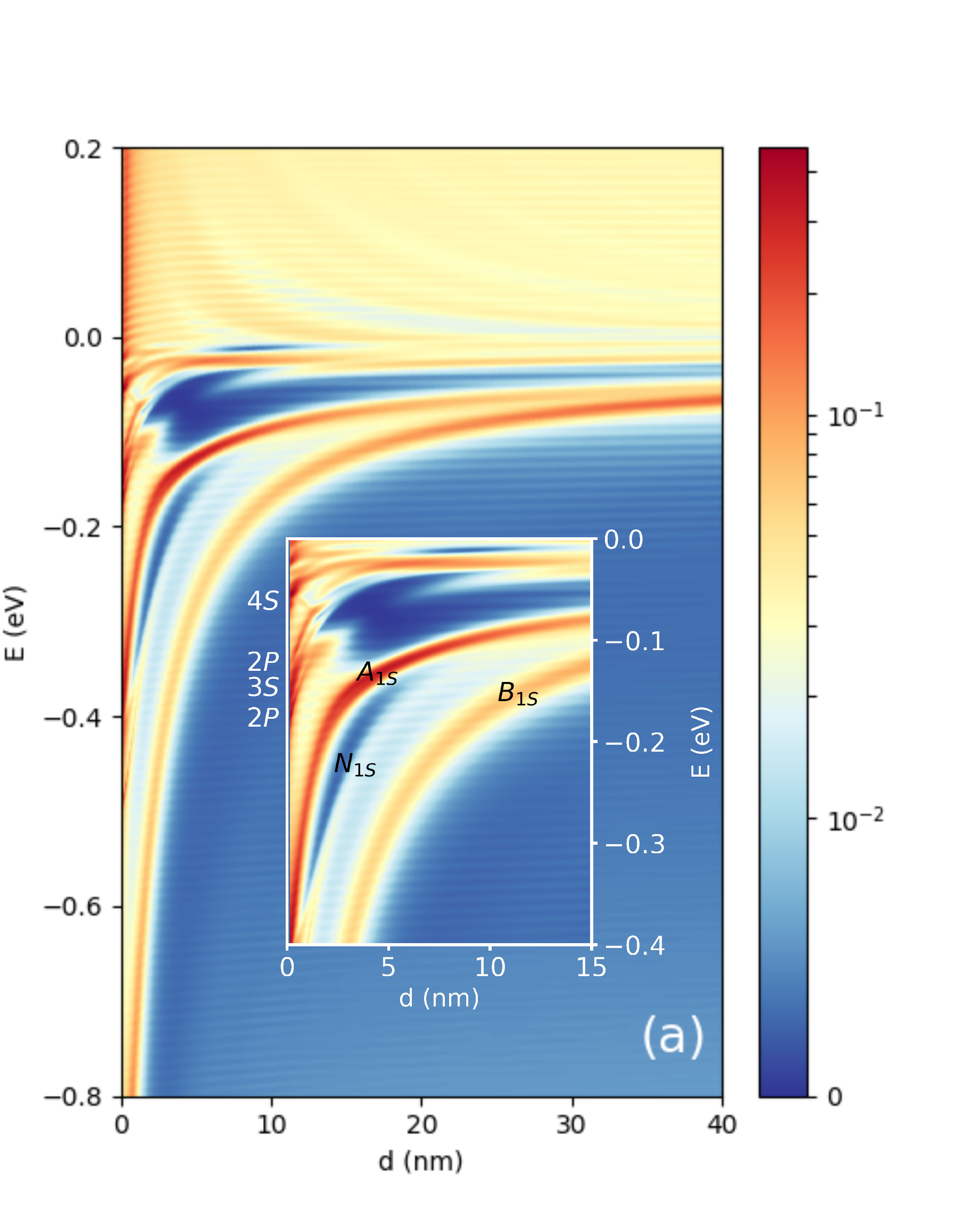}
\includegraphics[scale=0.47]{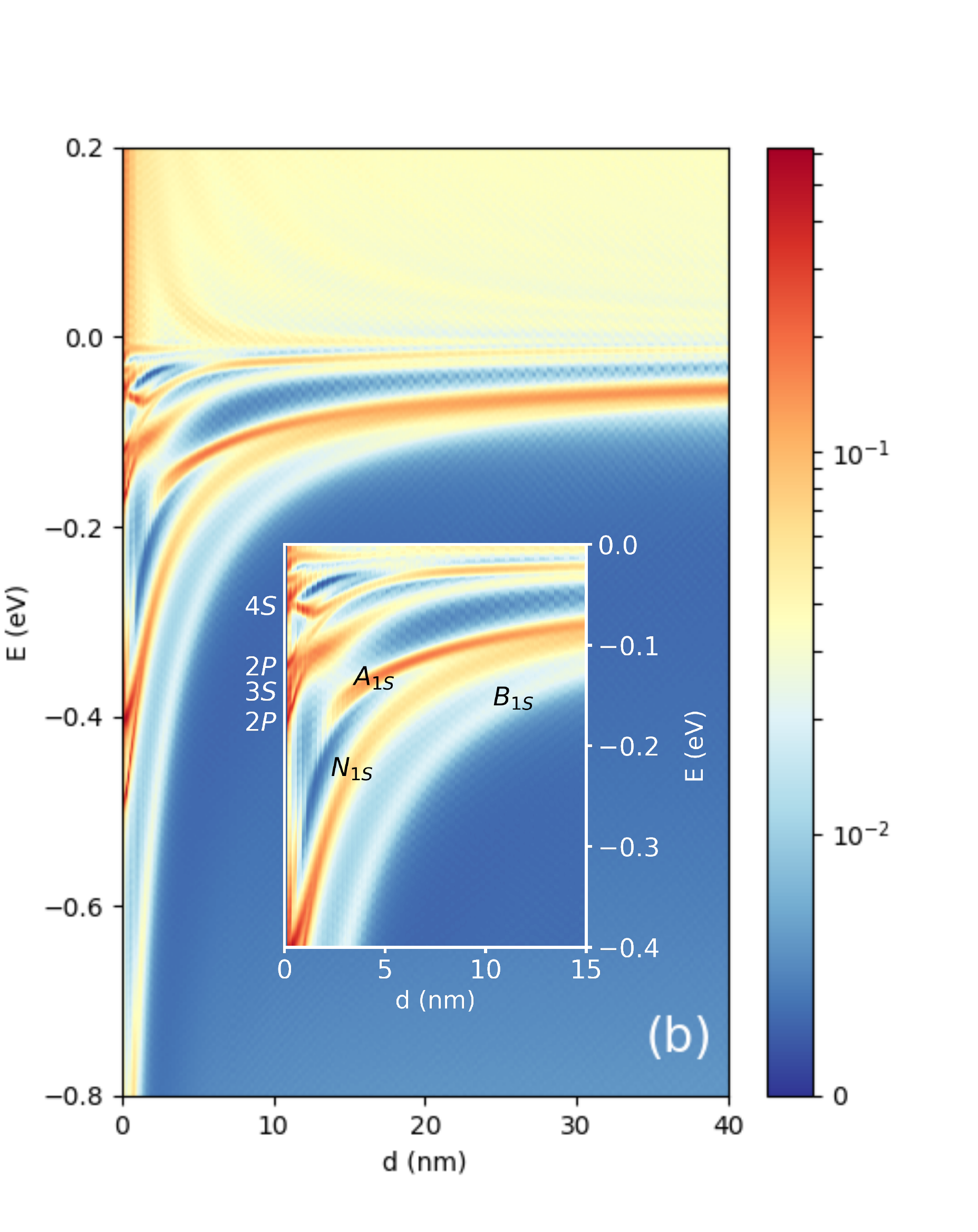}
\caption{LDOS for a system consisting of three charges of equal strength ($\beta=1$) arranged in a linear configuration. In (a) the LDOS is calculated at the position of the middle charge while for (b) the LDOS is calculated at one of the outer charges. In the insets a close up of the molecular collapse resonances, which are labeled in black, are shown.}
\end{figure*}
3) \textit{Molecular collapse region}: in this region both charges are individually supercritical and support an atomic collapse resonance. In this region the single impurity resonances interact and form molecular collapse resonances akin molecular orbitals from atomic physics. The characteristic length scale is determined by the regularisation distance $r_0$, which is reflected by a sharp peak in the LDOS at the impurity sites. The molecular collapse resonances converge towards the single impurity resonances with increasing charge and distance instead of being quenched.


\section{Three-charge system}
It is an interesting question to ask how the spectrum changes with increasing number of individual charges. Here we will extend the results of the previous section to a system consisting of three charges equal in strength placed in a linear configuration. This is a natural configuration, showing rich physics, that can be experimentally realized in an array of charges, as demonstrated in Ref. [\onlinecite{Crommie1}]. We will show that similarly to the two-charge system the resonance splits up into a molecular bonding and anti-bonding orbital. However, due to the odd number of charges an additional molecular resonance appears which we show to be the unstable counterpart of a \textit{non-bonding} molecular orbital. This system is an analog of the $H_3$ molecule [\onlinecite{Kimball}]. 

The three charge system is modelled by the following potential:

\begin{equation}
V_3(x,y)=V(x-d,y)+V(x,y)+V(x+d,y).
\end{equation}

In Fig. 9 we show the LDOS calculated for three charges of equal strength in a linear configuration as function of the energy and distance $d$ between the charges. In Fig. 9(a) the LDOS is calculated at the center charge while in Fig. 9(b) the LDOS is calculated at one of the outer charges. For large inter-charge distances a clear and distinct resonance is visible corresponding to the single impurity atomic collapse resonance $1S$ which is the unstable counterpart of the $1S$ atomic orbital. With decreasing inter-charge distance this resonance starts to split. However, in contrast with the two-charge system three instead of two distinct resonances emerge with decreasing inter-charge distance. This splitting into three resonances is more distinct when calculating the LDOS at one of the outer charges, see Fig. 9(b). 

\begin{figure}
\includegraphics[scale=0.40]{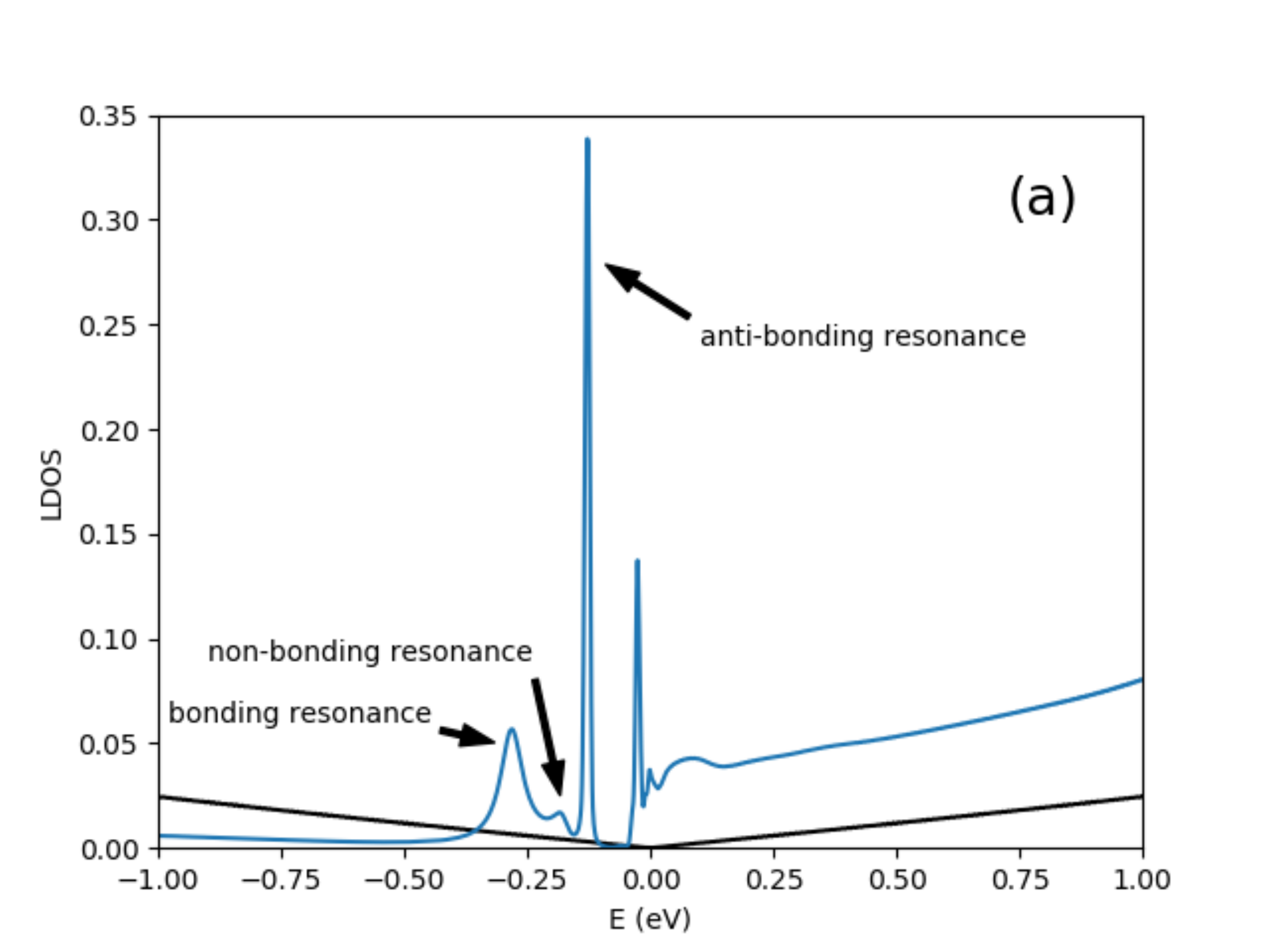}
\includegraphics[scale=0.40]{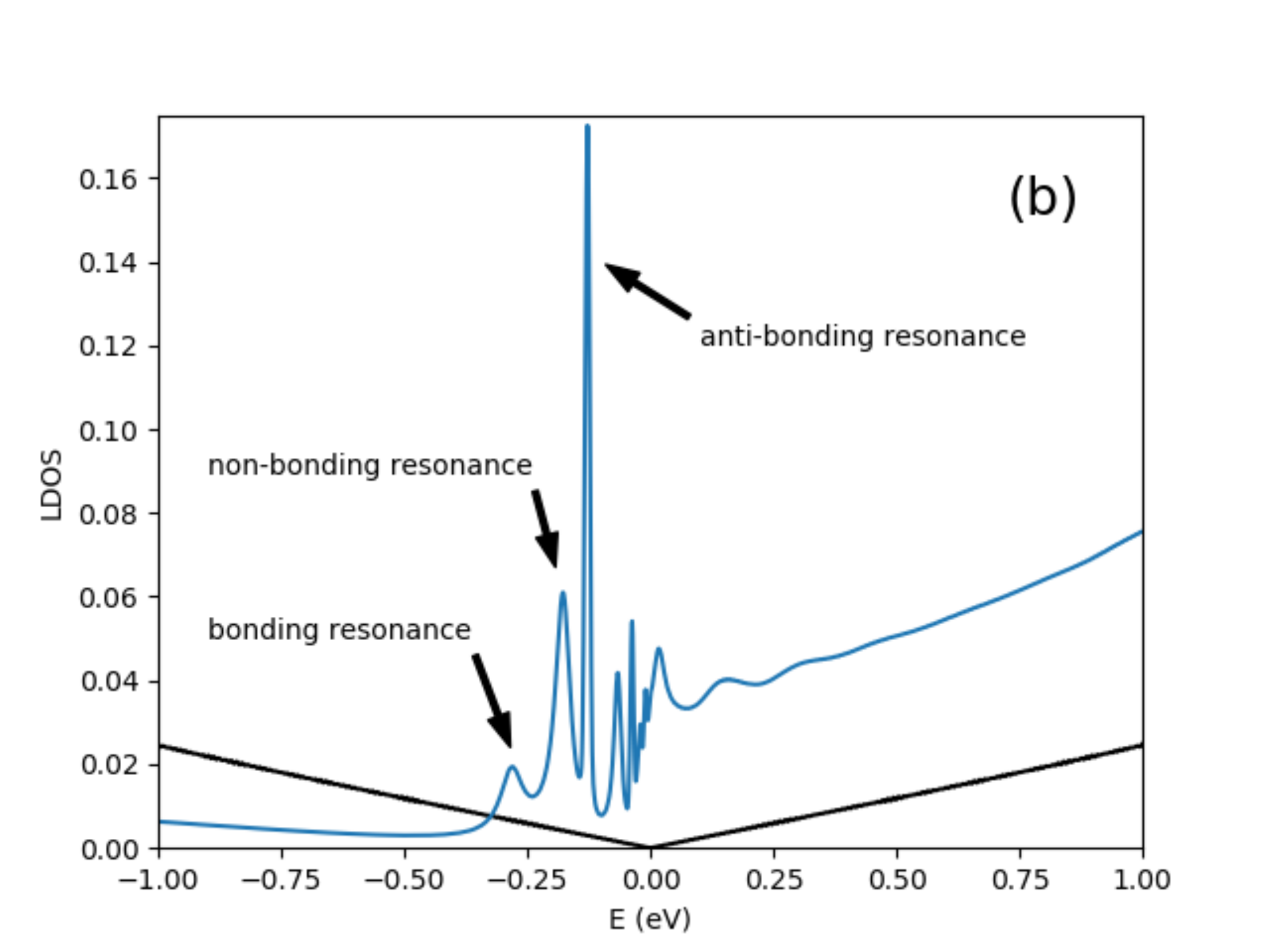}
\caption{(a) LDOS as function of the energy for three charges at a distance $d=5$ nm from each other. The LDOS is measured at the central charge as in Fig. 8(a). (b) The same as in (a) but this time the LDOS is calculated at one of the outer charges as in Fig. 8(b). The black curve is the LDOS of pristine graphene. A slightly larger flake size of $300$ nm was used in the calculation.}
\end{figure}

When calculating the LDOS at the center charge the single impurity resonance, visible for larger distances, seems to split up in only two distinct resonances (labeled as $B_{1S}$ and $A_{1S}$), see Fig. 9(a).  This result is very similar to the two-charge results shown in the previous section: the lowest energy branch has a lower intensity consistent with a molecular bonding orbital, the higher energy branch has clearly a higher intensity at the impurity which is consistent with an anti-bonding orbital. The resonance between those two peaks (labeled as $N_{1S}$) clearly behaves in a distinct and more special way: this resonance is only visible when calculating the LDOS at one of the outer charges. 

In order to understand the above behavior we look at the text book example of a linear $H_3$ molecule within the LCAO method. In this method one can use the single impurity orbitals in order to estimate the spatial distribution and energy of the molecular orbitals. In the LCAO method one can use the $1S$ single impurity orbitals of the individual charges, which we denote as $\ket{1S_a}$ and $\ket{1S_c}$ for the outer charges and $\ket{1S_b}$ for the middle charge, and use their linear combination as a trial wave function. It can be shown that using the three single impurity orbitals the following three molecular orbitals are obtained [\onlinecite{Borshch}]:
\begin{equation}
\ket{B_{1S}}=\frac{1}{2}\left(\ket{1S_a}+\sqrt{2}\ket{1S_b}+\ket{1S_c}\right),
\end{equation}
\begin{equation}
\ket{N_{1S}}=\sqrt{\frac{1}{2}}\left(\ket{1S_a}-\ket{1S_c}\right),
\end{equation}
\begin{equation}
\ket{A_{1S}}=\frac{1}{2}\left(\ket{1S_a}-\sqrt{2}\ket{1S_b}+\ket{1S_c}\right).
\end{equation}
The molecular orbital $\ket{B_{1S}}$ is a linear combination resulting in a chemical bond, hence the name \textit{bonding} orbital. The molecular orbital $\ket{A_{1S}}$ works against the bond and is called an \textit{anti-bonding} orbital. However, their is also a third molecular orbital $\ket{N_{1S}}$ which has no contribution from the wave function located at the central charge. This orbital has both bonding and anti-bonding character and is therefore called a \textit{non-bonding} orbital. The correspondence of our results with the results obtained within the LCAO method are remarkable: the middle resonance visible in Fig. 9 behaves exactly like the non-bonding molecular orbital of the $H_3$ molecule which is reflected by the fact that this resonance is not visible when measuring the LDOS at the central impurity. 

In Fig. 10 we present a cut of the LDOS of Fig. 9 at a fixed inter-charge distance $d=5$ nm. In Fig. 10(a) the LDOS is calculated at the central charge, as described in the previous paragraph the peaks of the anti-bonding and bonding molecular resonances should present a clear signature to look for in experiments. However, when calculating the LDOS at one of the outer charges as in Fig. 10(b) the non-bonding resonance should appear between the bonding and anti-bonding molecular resonances in an experiment.  
\begin{figure}
\includegraphics[scale=0.43]{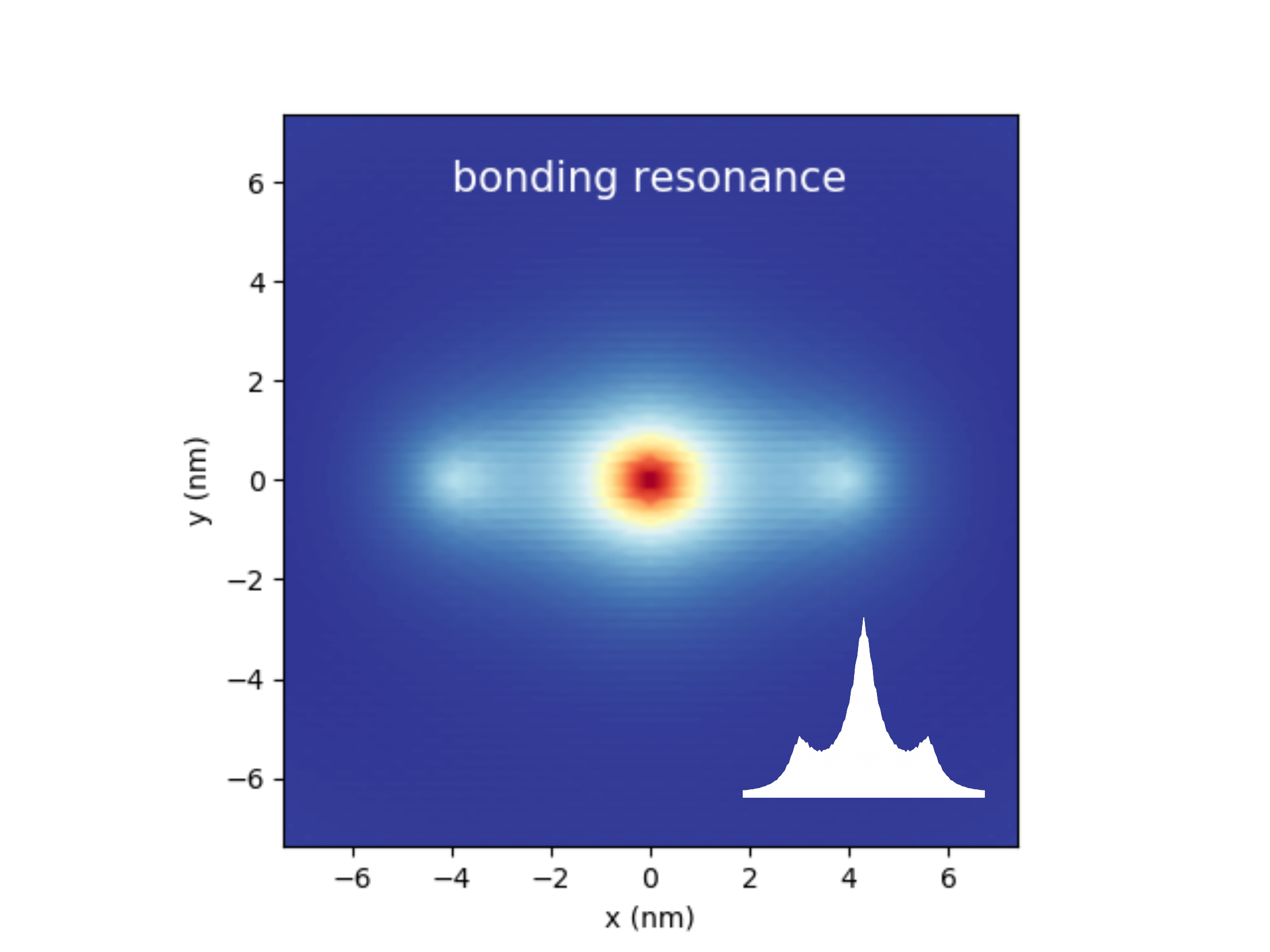}
\includegraphics[scale=0.43]{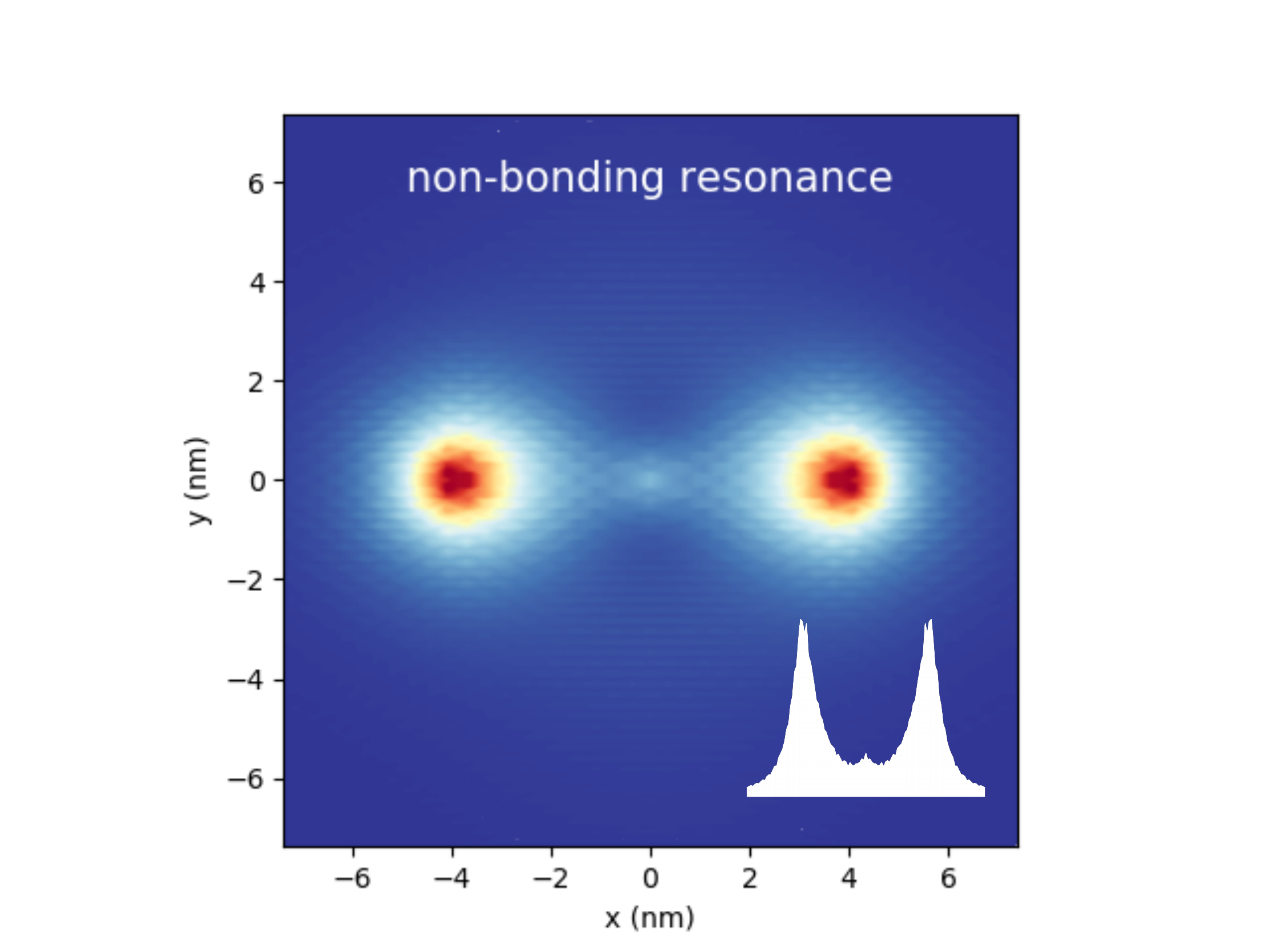}
\includegraphics[scale=0.43]{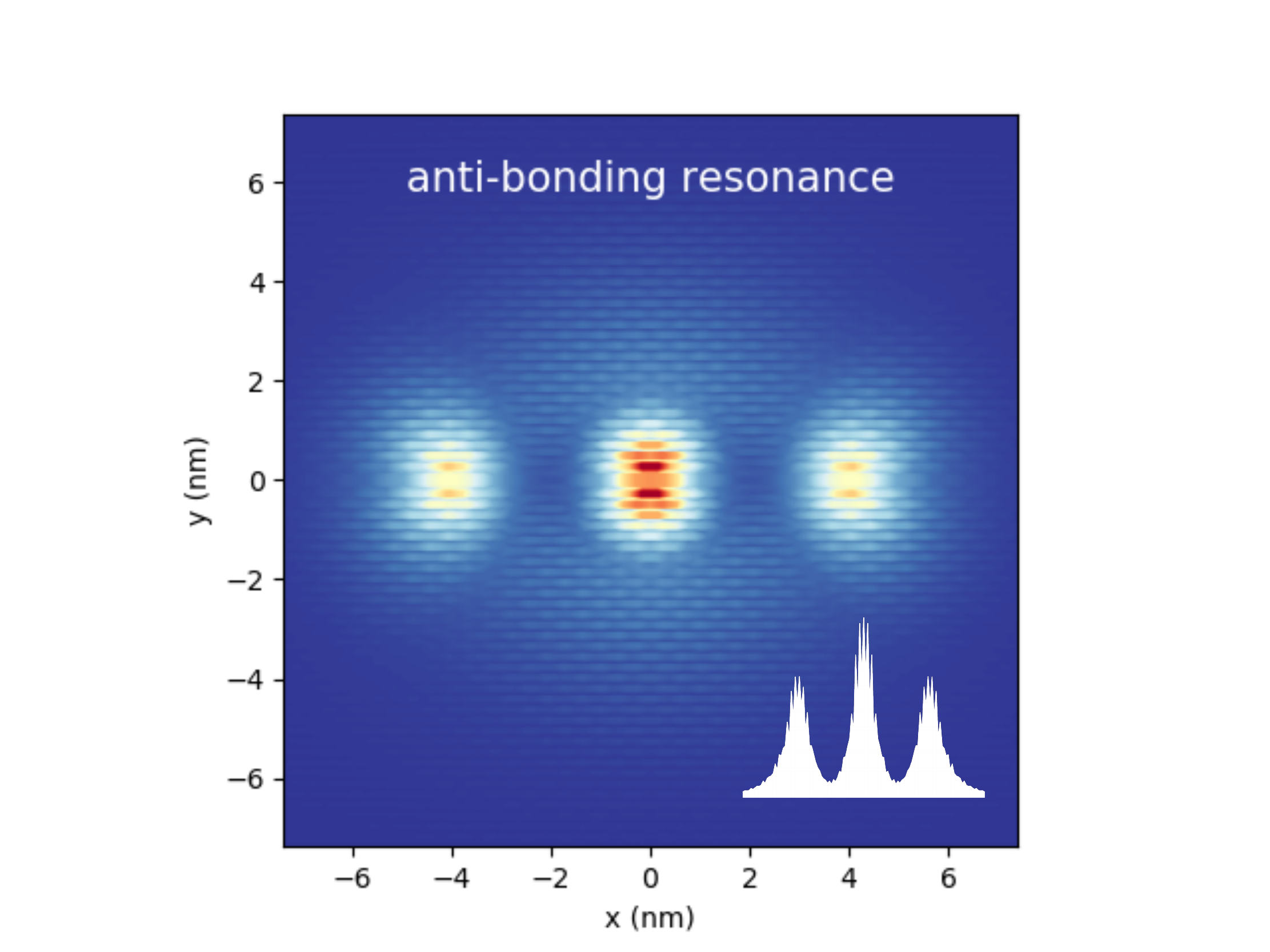}
\centering
\caption{Countour plot of the spatial LDOS for the bonding, non-bonding and anti-bonding resonance, respectively, for three impurities separated at a distance $d=4$ nm from each other. The spatial LDOS is calculated for the energies $E=-0.35$ eV (bonding: $B_{1S}$), $E=-0.2$ eV(non-bonding: $N_{1S}$) and $E=-0.15$ eV (anti-bonding: $A_{1S}$).}
\end{figure}

In order to support the statements made in the previous paragraph we investigate the spatial LDOS for the three resonances observed in Fig. 9. This is presented in Fig. 11 where the spatial LDOS is plotted for the bonding, anti-bonding and non-bonding resonances for the three charges separated at a distance $d=4$ nm from each other and each with individual charge of $\beta=1$. The bonding orbital clearly has a finite LDOS between the impurities enhancing the bond between the charges, exactly as in the two-charge case which is typical for a bonding molecular state. Note how the intensity of the spatial LDOS is larger for the central charge, this behaviour exactly corresponds to the prediction made by the LCAO method (see Eq. (5)). For the non-bonding orbital the spatial LDOS has almost no intensity on the middle charge while it has equal density on the two outer charges. This is exactly the kind of behaviour expected for the non-bonding orbital described by the LCAO method, see Eq. (6). Last we calculated the spatial LDOS for the anti-bonding resonance. Similar to the two-charge case a very low spatial LDOS value between the impurities is found working against the bond which is typical for anti-bonding orbitals. 

From the previous results it was shown that the molecular collapse resonances behave like the molecular orbitals in a linear $H_3$ molecule. This confirms that the molecular collapse resonances can be considered as the unstable counterparts of molecular orbitals in molecular physics. 


One important remark should be made. While there is very good qualitative agreement between the LCAO results for the $H_3$ and our results for the molecular collapse resonances, quantitatively deviations are observed. This can be explained by the fact that: i) the LCAO method is an approximation, ii) we are working with quasi-bound states instead of bound states, and iii) the linear spectrum of graphene. 


We also calculated the LDOS for three charges in a symmetric triangular configuration. In such a system calculating the LDOS on the impurity charge gives the same result for all the charges due to the symmetric nature of the configuration. We found that the behavior of the resonances as function of the inter-charge distance is qualitatively the same: the single impurity resonance splits up into three new molecular collapse resonances.

\section{Conclusions}

In this paper, we studied how the atomic collapse effect manifests itself in the presence of multiple charges.  This work is substantially different from previous studies with multiple charges where the existence of the collapse states was investigated versus the number of sub-critical charges [\onlinecite{Saffa}] or their distance and charge [\onlinecite{Saffa}-\onlinecite{Egger}]. Here, we extended these results to charges that are individually supercritical and showed that they form \textit{new kind of resonances} which mimic molecular orbitals known from atomic physics. We showed the emergency of three distinc collapse regions in the presence of multiple charges: subcritical region, frustrated atomic collapse region and molecular collapse region. We pointed out the main differences between the different regions and compared our results for the LDOS and spatial distribution of the LDOS for the different regions. 

Systems consisting of two and three charges in a linear spatial configuration were investigated. For the two-charge system we showed that the single impurity resonance splits up into two resonances, one of which is lower in energy and has a bonding character reflected both in the LDOS and the spatial LDOS while the other is higher in energy and has a clear anti-bonding behaviour. In the case of three charges the behaviour is different. The single impurity resonance splits up into three resonances. The lowest energy resonance has a clear bonding character and the highest energy resonance has an anti-bonding character. For both the bonding and anti-bonding resonance the spatial LDOS showed a peak at the central impurity. However, between the bonding and anti-bonding molecular resonances another resonance is found which behaves differently from the other two. The spatial LDOS shows almost no intensity at the central impurity and equal density on the two outer charges. On top of that the resonance has a mixed bonding and anti-bonding character.

A phase diagram was constructed that demarcates the three different and distinct regions: the subcritical region, frustrated atomic collapse region and molecular collapse region. In the subcritical region no atomic collapse resonances are visible, regardless of the inter-charge distance. In the frustrated atomic collapse region an atomic collapse resonance is visible, this resonance has a spatial distribution of the order of the inter-charge distance and becomes quenched with increasing inter-charge distance. In the molecular collapse region single impurity resonances interact with each other and form molecular resonances akin molecular orbitals from atomic physics. In the frustrated atomic collapse region the mechanism inducing atomic collapse is very similar to that of the single impurity case but with the regularization distance replaced by the inter-charge distance, while in the molecular collapse region single impurity resonances interact and form molecular collapse resonances with a characteristic spatial dependance.  


In conclusion we showed that the atomic collapse resonances induced by supercritical charges interact and form new kind of resonances, called molecular collapse resonances. Our results show that graphene with supercritical charges could provide a platform to detect and characterize such molecular orbitals. Next to that we showed that by changing the inter-charge distance, the individual impurity strength and the number of charges a high degree of control over the molecular collapse states can be achieved. Our predictions are experimentally detectable by using multiple supercritical charge clusters and measuring the LDOS using an STM tip. The theoretical model presented in this manuscript has shown good agreement between theory and experiment in earlier single impurity atomic collapse systems [\onlinecite{Crommie}, \onlinecite{Peeters}]. The molecular collapse resonances should show clear peaks in the LDOS when measured at one of the impurity charges. The spatial LDOS of these molecular collapse resonances should show a clear bonding, anti-bonding or non-bonding structure in their spatial distribution. 

\acknowledgements

We thank Matthias Van der Donck for fruitful discussions. This work was supported by the Research Foundation of Flanders (FWO-V1) through an aspirant research grant for RVP and a postdoctoral grant for SPM. 
\newline

\appendix

\section{Resonances for $E>0$}

\begin{figure}[h!!!!]
\includegraphics[scale=0.45]{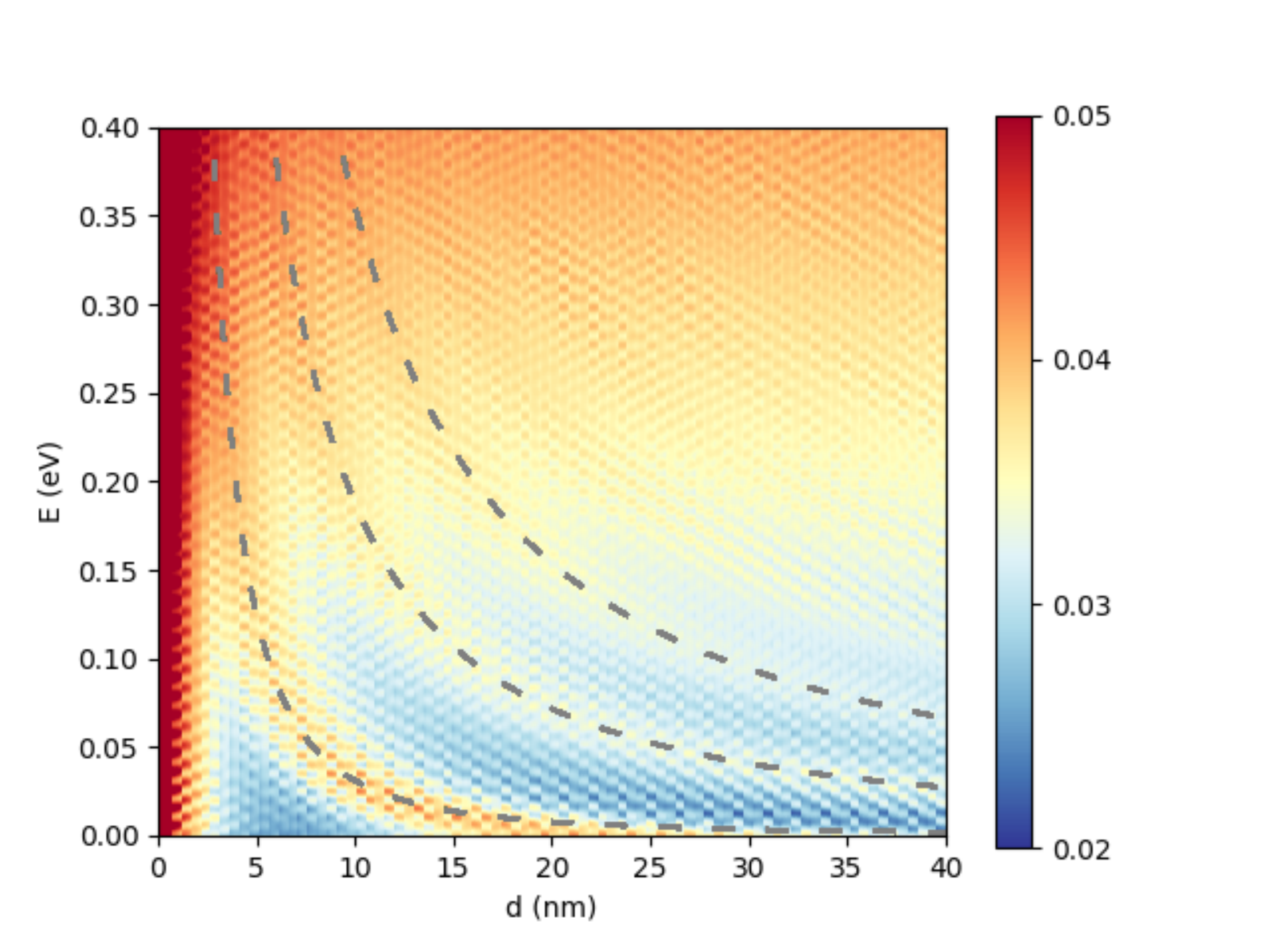}
\caption{Close up of Fig. 2(a) of the positive energy region for which weak resonances are observed. The energy scale is chosen such that the low intensity resonances are enhanced.}
\end{figure}

In Fig. 2(a) weak resonances are observed for positive energy. These resonance are broad and have low intensity. In Fig. 12 a close up of the positive energy range of Fig. 2(a) is shown where the scale for the LDOS is chosen in such a way that the broad and low intensity resonances are made more visible. In this close up, three distinct broad resonances can be observed. With decreasing inter-charge distance these resonances rise in energy. Given the different characterstics of these resonances and there opposite d-dependence as compared to the atomic collapse resonances suggest that they have a different origin. From their dependence on the inter-charge distance we can suspect that they emerge from scattering between the two charges causing interference between the electron scattering states. This interpretation is supported by the spatial LDOS shown in Fig. 13. In Fig. 13(a) the spatial LDOS is shown for a point belonging to the lowest energy resonance shown in Fig. 12 for $d=10$ nm. The spatial LDOS exhibits interference between the two charges. In Fig. 13(b) the spatial LDOS is shown for the second resonance shown in Fig. 12 for $d=10$ nm. Again a clear spatial structure between the two charges is observed consistent with interference caused by the two charges. 

We fitted the functional behavior of the resonances to the function $E(d)=a.d^b$, with $a$ in units of $eV$ and $d$ in nm, and found respectively the values $\{a,b\}=\{3.11, -2\},\text{ }\{4.64, -1.39\}, \text{ } \{5.76,-1.21\}$ for the lowest three resonances which are shown as dashed curves in Fig. 12. For the lowest energy resonance the behavior is almost perfectly $E\propto 1/d^2$. This is exactly the kind of behavior expected for resonances in a square potential barrier for energies $E>V$ [\onlinecite{Masir}]. Since the two charges act as the edges of a potential well creating standing waves in between. For higher energy resonances the behaviour deviates from that of a perfect square well due to differences in the geometric structure of our system and the latter. 

\begin{figure}
\includegraphics[scale=0.43]{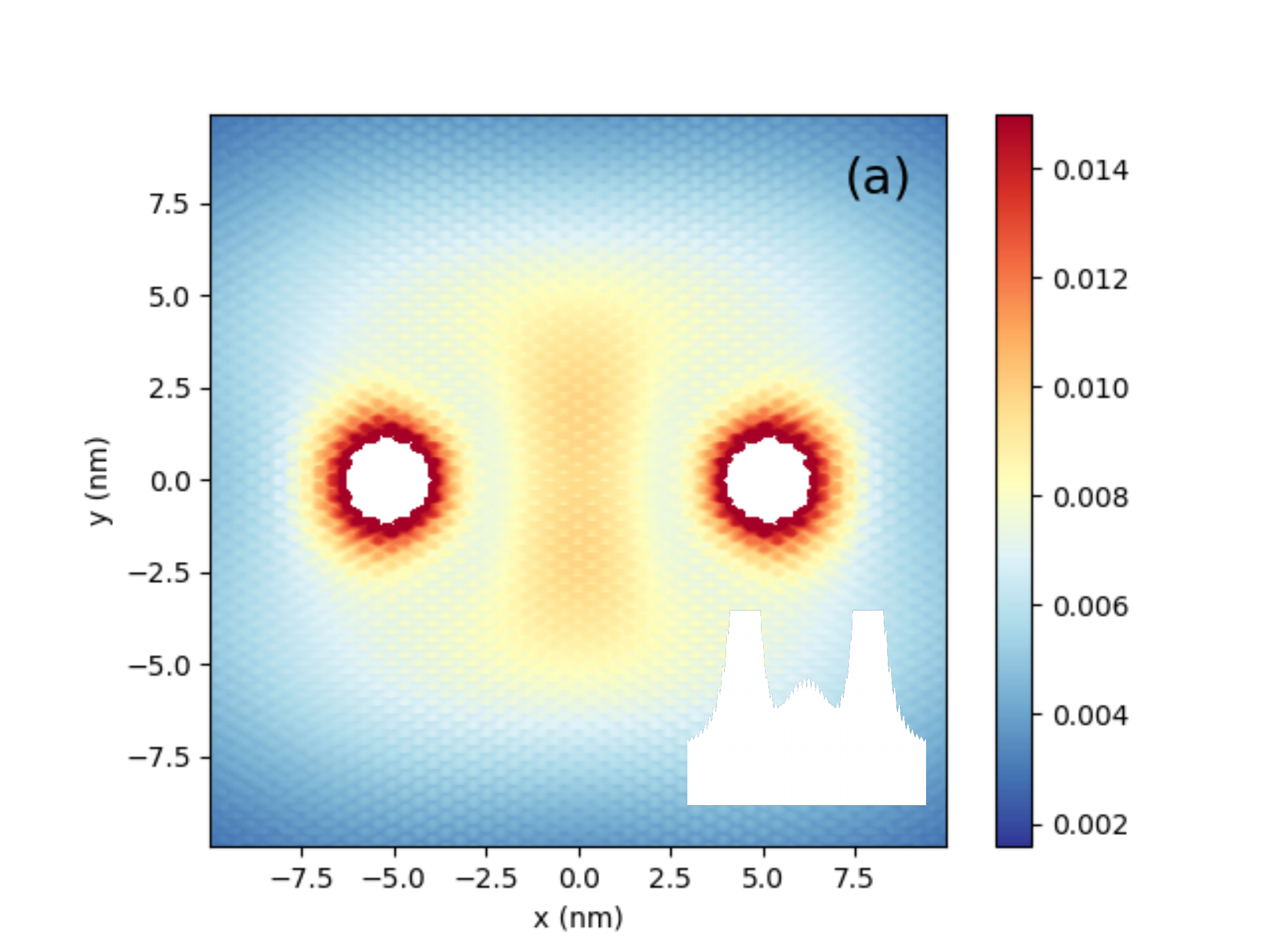}
\includegraphics[scale=0.43]{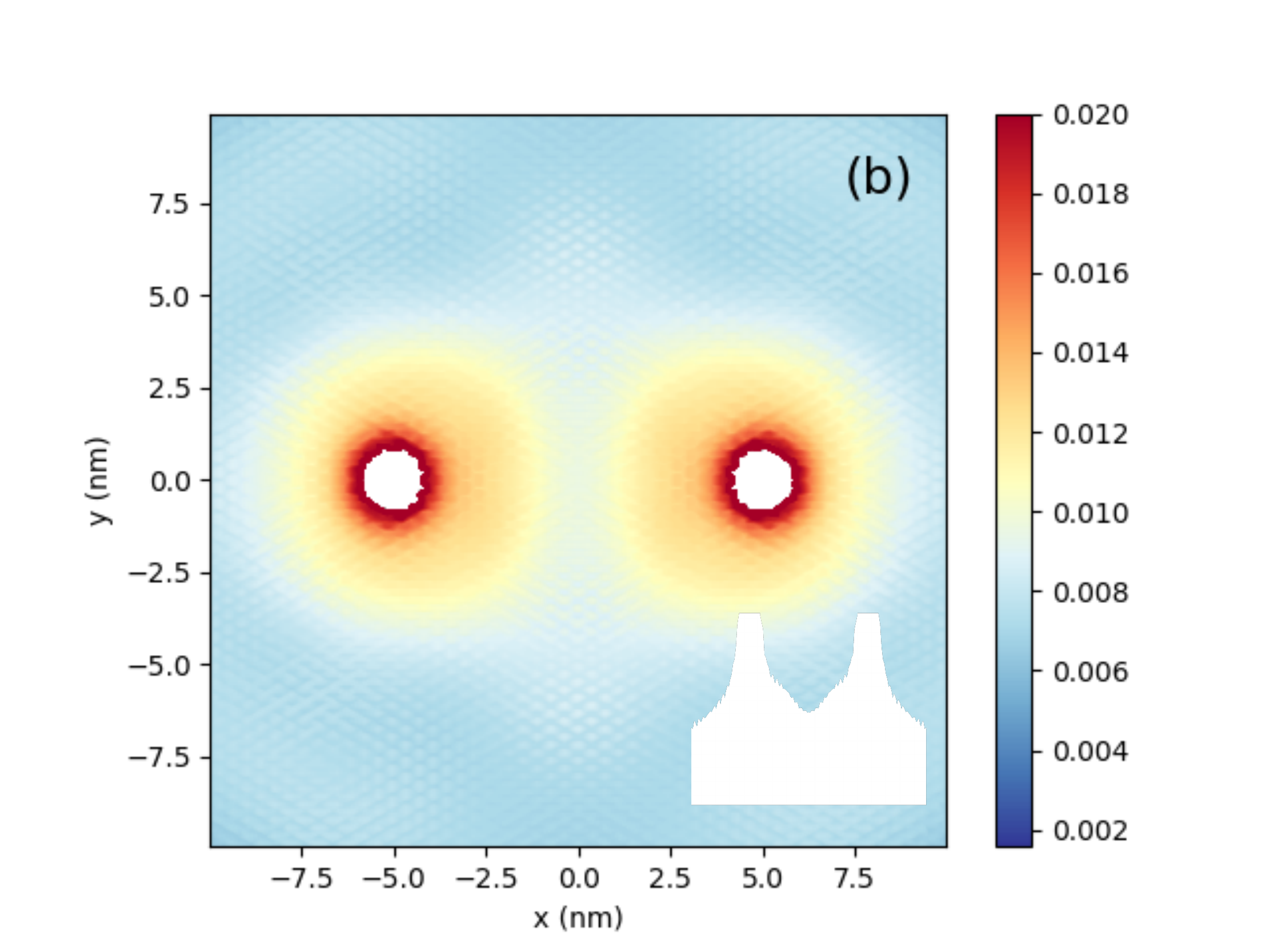}
\caption{(a) Spatial LDOS measured at an energy $E=0.04$ eV for two charges at a distance $d=10$ nm from each other. (b) The same as (a) but for $E=0.19$ eV. The LDOS near the impurity is truncated (white circular region) in order to enhance the visibility of the LDOS between the impurities.}
\end{figure}

\end{document}